\pdfoutput=1
\documentclass[PAPER, atlasdraft=false, UKenglish, cernpreprint, texmf, orcidlogo]{atlasdoc}
\usepackage{atlaspackage}
\usepackage{multirow}
\usepackage{bm}
\usepackage{siunitx}
\usepackage{atlasbiblatex}

\usepackage{atlasphysics}
\graphicspath{{logos/}{figures/}}

\usepackage{ANA-EXOT-2018-49-PAPER-defs}

\AtlasTitle{Search for $t\bar{t}$ resonances in final states with exactly one or two leptons using 140~fb$^{-1}$ of \(pp\) collision data at \sqrtS with the ATLAS experiment}

\AtlasAbstract{%
A search for heavy spin-1 and spin-2 resonances decaying into a top-antitop-quark pair has been performed with 140~fb$^{-1}$ of proton--proton collision data collected by the ATLAS experiment at the Large Hadron Collider at a centre-of-mass energy of \sqrtS. Final states with either exactly one electron or muon, or exactly two leptons (\(ee\), \(\mu\mu\) or \(e\mu\)), large missing transverse momentum, and two jets, at least one of which must be identified as likely containing a \textit{b}-hadron decay, are considered. The search targets resonances with both narrow and broad widths relative to the detector resolution, and with masses in the range of 0.4--5.0~TeV. No significant deviation from the Standard Model prediction is observed. Exclusion limits are set on the production cross-section times branching ratio for hypothetical $Z'$ bosons, Kaluza–Klein gravitons, and Kaluza–Klein gluons that decay into top-quark pairs.
}

\AtlasRefCode{EXOT-2018-49}

\PreprintIdNumber{CERN-EP-2025-281}

\AtlasJournal{JHEP}
\AtlasJournalRef{\JHEP 10 (2026) 219}
\AtlasDOI{10.1007/JHEP10(2026)219}

\hypersetup{pdftitle={ATLAS document},pdfauthor={The ATLAS Collaboration}}

\begin{document}

\maketitle

\tableofcontents

\section{Introduction}
%
%

New heavy particles coupling strongly to the top quark are predicted in numerous extensions of the Standard Model (SM) of particle physics. Such models are proposed to address open questions of the SM, such as the large difference between the scales of electroweak interactions at \(\mathcal{O}(100)\)~\GeV and gravity at \(\mathcal{O}(10^{19})\)~\GeV, known as the Planck scale~\cite{Hooft1980,Veltman:1980mj}. Models include, for example, topcolor-assisted-technicolor (TC2)~\cite{Hill:1993hs,Hill:1994hp,Harris1999,Harris:2011ez} and Randall--Sundrum (RS) models of warped extra dimensions~\cite{Randall:1999ee, Lillie:2007yh}, predicting one or several additional heavy bosons, such as a spin-1 \zprime particle, a spin-2 Kaluza--Klein (KK) graviton (\grav), or a spin-1 KK gluon (\glu). The new states can have strong or preferential couplings to the top quark, the heaviest elementary particle of the SM. Resonant production of a top--antitop-quark pair (\ttbar), if observed, would thus provide a smoking-gun signature of phenomena beyond the SM. Searches for this process are therefore an integral part of the experimental programme at the Large Hadron Collider (LHC)~\cite{Evans:2008zzb}, which now provides unique access to heavy resonances with masses in the multi-\TeV range.

Searches for heavy resonances decaying into a \ttbar pair have previously been conducted with the ATLAS and CMS experiments using proton--proton (\(pp\)) collisions at \sqrtSseven~\cite{TOPQ-2011-12,TOPQ-2011-23,TOPQ-2012-14,CMS-EXO-11-006,CMS-TOP-11-010,CMS-TOP-12-017}, 8~\TeV~\cite{CMS-B2G-13-001,TOPQ-2012-18,CMS-B2G-13-008,EXOT-2016-04}, and 13~\TeV~\cite{EXOT-2015-04, EXOT-2016-24, CMS-B2G-16-015, CMS-B2G-17-017, EXOT-2018-48}. Both experiments have performed the searches in the fully-hadronic and single-lepton final states, while the CMS Collaboration has also included the dileptonic final states. ATLAS and CMS have also recently performed dedicated searches for (pseudo-)scalar resonances~\cite{EXOT-2020-25,CMS-HIG-22-013}, focusing on signals with masses below 2~\TeV and where interference between the signal and the SM \ttbar background is known to be significant. These signals are hence not considered in this paper.  In the search performed by CMS, a significant deviation from the SM expected limits for a \ttbar mass less than approximately 400~\GeV was observed. This was later shown by CMS~\cite{CMS-TOP-24-007} to be consistent with a colour-singlet, quasi-bound state of top quarks, predicted in the SM. The mild tension observed by ATLAS~\cite{TOPQ-2021-24} and CMS~\cite{CMS-TOP-23-001} between the data and simulation in the quantum entanglement studies is also partially relieved by the inclusion of the pseudo-scalar state. In this paper, the focus is on searches for resonant signals with masses above 400~\GeV; the inclusion of the quasi-bound state as a background was tested and has no impact on the results of this paper.

This paper presents a search for \ttbar resonances across the \ttbar invariant mass spectrum, \mttbar, from 0.4~\TeV to 5.0~\TeV. The search is performed using 140~\ifb of \(pp\) collision data recorded in 2015--2018 at \sqrtS with the ATLAS detector at the LHC, and uses the single-lepton and dilepton final states, complementing the fully-hadronic ATLAS search. The results in this paper consider spin-1 and spin-2 signal models and are valid for the case when interference is negligible. It is the first time in Run 2 that the dilepton channel has been included in an ATLAS search for these types of resonances, and both the single-lepton and dilepton channels use the full Run-2 data for the first time. Additionally, the single-lepton channel improves on the previous 36~\ifb analysis through improved \ttbar reconstruction and the addition of validation regions to better evaluate the background modelling.

The search presented in this paper targets events in which one top quark decays semi-leptonically and the other hadronically (1-lepton channel), as well as events in which both top quarks decay leptonically (2-lepton channel). Here, leptonic top-quark decays are defined as the top-quark decays resulting in a prompt electron (\(e\)) or muon (\(\mu\)), either directly from the \textit{W} boson decay or via secondary decays of leptonically-decaying \(\tau\)-leptons. Hadronic top-quark decays in the 1-lepton channel are targeted using separate analysis strategies for resolved and merged topologies that are combined to maximise the sensitivity for the \mttbar range from 400~\GeV up to values of several \TeV. In the range \(m_{t\bar{t}}>2\)~\TeV, the top quarks are highly Lorentz boosted, resulting in the decay products merging into a single jet. Therefore, an advanced multivariate classifier based on a deep neural network is used to identify the hadronic top-quark decays with these highly collimated decay products~\cite{JETM-2018-03}. The dominant background in both channels arises from SM \ttbar production.

This paper is structured as follows. In Section~\ref{sec:models}, the benchmark models used for the interpretation of the results of this search are introduced. In Sections~\ref{sec:detector} and~\ref{sec:datamc}, the ATLAS detector as well as the data and simulated event samples are described. The object reconstruction, event selection, categorisation and reconstruction of observables, and the estimate of background processes are described in Sections \ref{sec:objects} to \ref{sec:bkg}. The description of the statistical analysis is provided in Section~\ref{sec:stats} and the systematic uncertainties are detailed in Section~\ref{sec:unc}. Finally, results are presented and discussed in Section~\ref{sec:results}, where, in the absence of any significant excess, exclusion limits on the production cross-section times branching ratio for the hypothetical new particles are presented at the 95\% confidence level.


%
\section{Theoretical framework}
\label{sec:models}
%

The search presented in this paper is designed to be sensitive to the production of different types of spin-1 and spin-2, \textit{s}-channel resonances decaying to \ttbar. The selection efficiency and acceptance can differ depending on the properties of the new resonance, such as its spin, width, or production mode. Therefore, the sensitivity to three types of new particle was evaluated to quantify the performance of the search. The benchmark models adopted in this search are colour-singlet and colour-octet bosons with spin-1 or spin-2 and masses ranging from 400~\GeV to 5~\TeV. Two benchmark models with a narrow relative resonance width, \(\Gamma/m\), below the experimental resolution of \(\Gamma/m \approx \mathcal{O}(10\%)\) are used, in addition to a model predicting wide resonances with a relative width of \(\Gamma/m = 30\%\). Figure~\ref{fig:feyn} shows example Feynman diagrams for the three benchmark models considered, and the following section details the theoretical motivation for such models.
\begin{figure}[t]
\centering
\subfloat[]{
\includegraphics[width=0.3\textwidth]{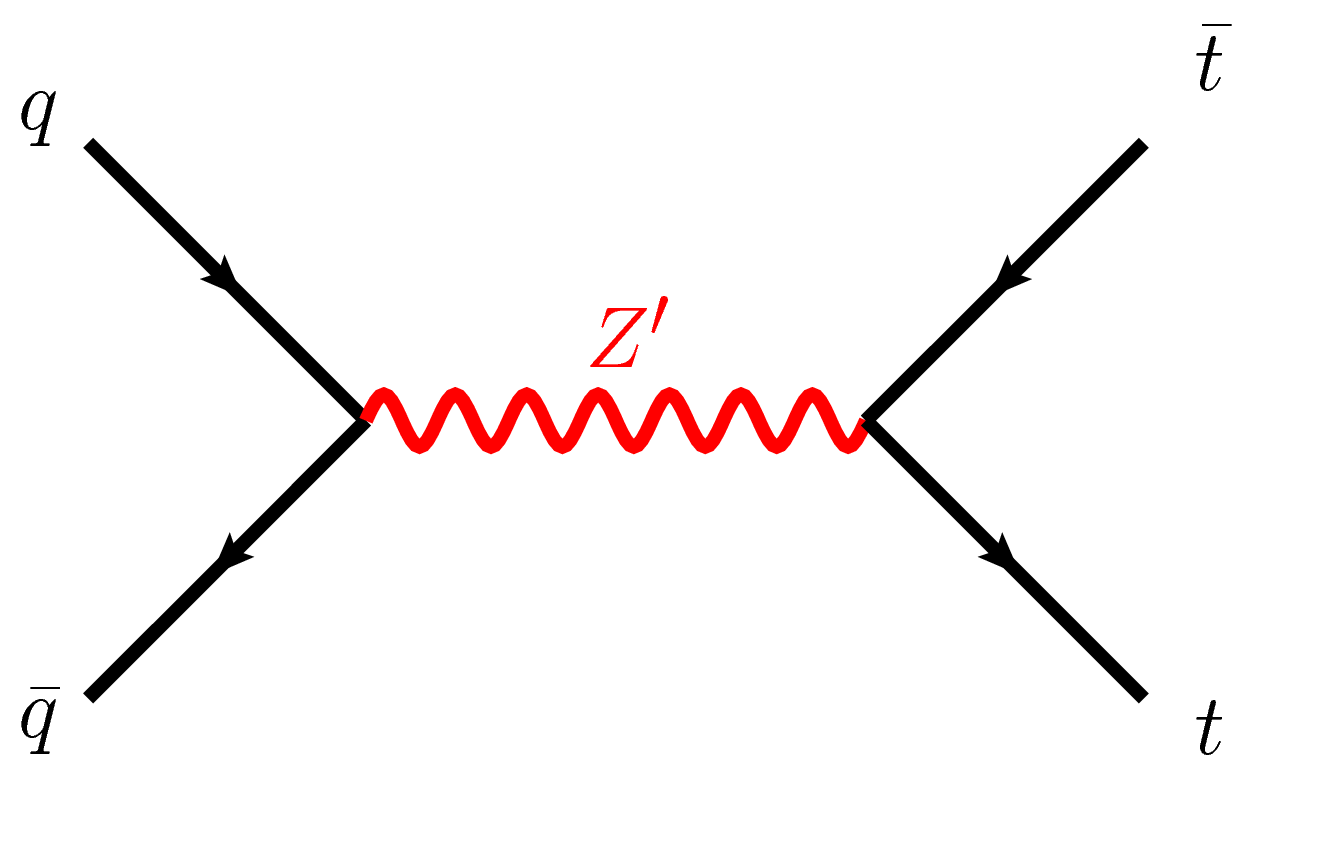}
}
\subfloat[]{
\includegraphics[width=0.3\textwidth]{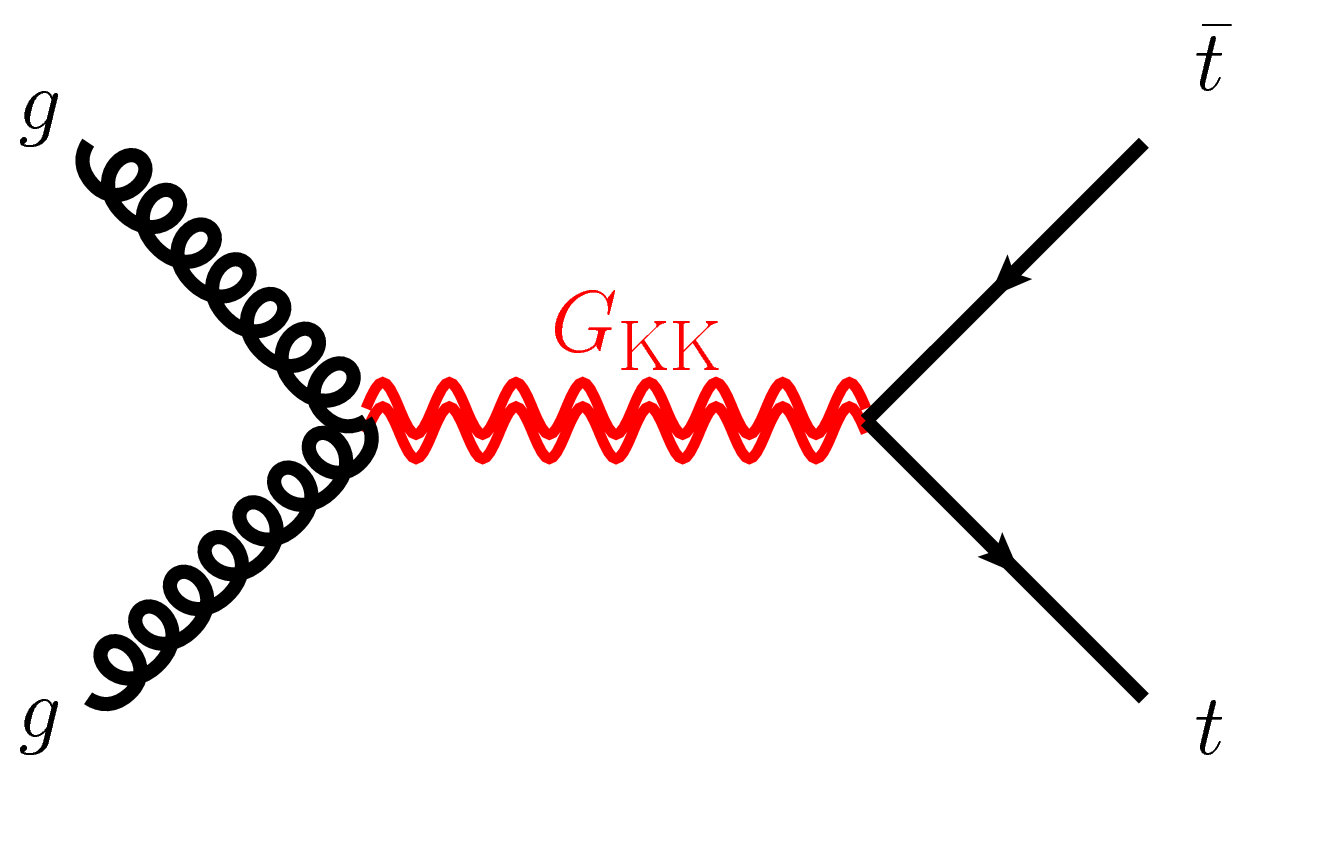}
}
\subfloat[]{
\includegraphics[width=0.3\textwidth]{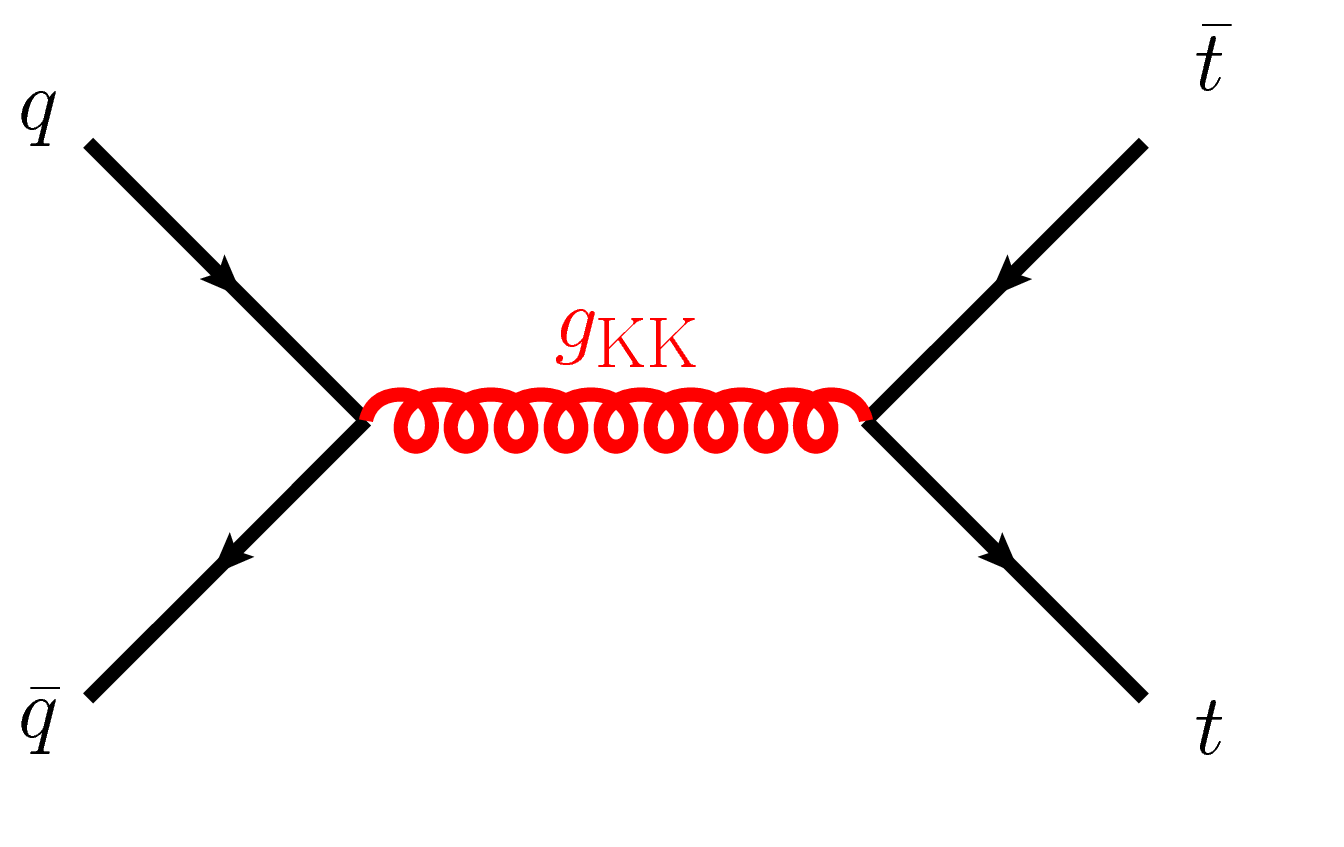}
}
\caption{
Representative leading-order Feynman diagrams for the \textit{s}-channel production and subsequent decay to \ttbar of the (a) \zprimeTC\,, (b) \grav, and (c) \glu signals.
\label{fig:feyn}}
\end{figure}

\subsection{Spin-1 colour singlet}
Spin-1, electrically neutral, colour-singlet vector bosons, referred to as \zprime bosons, are predicted in various extensions of the SM, such as topcolor-assisted-technicolour models~\cite{Hill:1993hs,Hill:1994hp}. The benchmark model used in this paper is the same as in previous searches for \ttbar resonances~\cite{EXOT-2018-48,EXOT-2015-04,EXOT-2016-24,TOPQ-2012-18,EXOT-2016-04,TOPQ-2011-23,TOPQ-2012-14} and is referred to as Model IV in Ref.~\cite{Harris1999}, with production cross-section calculations for the LHC provided by Ref.~\cite{Harris:2011ez}. The model predicts a leptophobic \zprime boson, referred to as \zprimeTC, which is mainly produced via \(q\bar{q}\) annihilation and decays into first- and third-generation quarks. The properties of the boson are controlled by three parameters: the top-color tilting parameter, \(\cot\theta_H\), which controls the width and the production cross-section, and \(f_1\) and \(f_2\), which are related to the coupling to up-type and down-type quarks, respectively. In the benchmark considered in this and previous papers, \(f_1\) is set to 1 and \(f_2\) to 0 to maximise the branching fraction of the \zptctt decay, which reaches 33\%. The parameter \(\cot\theta_{H}\) is tuned for each mass point such that the resonance has a width of 3\% of its mass~\cite{Harris:2011ez}. Given that this width is smaller than the experimental resolution, the \zprimeTC signal considered in this paper can also be used to set limits on the production cross-section times branching ratio for a \zprimeTC signal with other widths that are also below the experimental resolution by rescaling the theory production cross-section. The predicted next-to-leading-order (NLO) \zptctt production cross-section for a \zprimeTC mass of 4~\TeV is about 7.6~fb for a width of \(\Gamma/m=3\%\) and about 2.0~fb for a width of \(\Gamma/m=1\%\).

The strongest mass constraints on this model have been derived by the ATLAS Collaboration using the search in fully-hadronic final states on 139~\ifb of \sqrtS data~\cite{EXOT-2018-48}. This search excluded \zprimeTC bosons with 1\% and 3\% relative widths with masses up to 3.9~\TeV and 4.7~\TeV, respectively. The CMS Collaboration excluded \zprimeTC bosons with masses up to 3.80~\TeV for a decay width of 1\% using the combination of all three decay channels on 35.9~\ifb of \sqrtS data~\cite{CMS-B2G-17-017}. In the 1-lepton decay channel, the ATLAS Collaboration excluded \zprimeTC signals of width 1\% with masses up to 3.0~\TeV, using 36~\ifb of data~\cite{EXOT-2015-04} and no previous search in the 2-lepton decay channel has been performed by the ATLAS Collaboration. Dijet searches performed by ATLAS~\cite{EXOT-2019-03,EXOT-2018-09,EXOT-2019-01} and CMS~\cite{CMS-EXO-19-012} also place constraints on leptophobic \zprime bosons. ATLAS has provided direct sensitivity comparisons in the context of a simplified model with a dark matter \zprime mediator that has flavour-universal couplings to all quarks, and this shows similar sensitivities between flavour-inclusive dijet searches and \ttbar searches at high masses~\cite{EXOT-2018-62}. Given that the \zprimeTC model considered in this paper couples preferentially to top quarks, leading to approximately twice the branching ratio as that to bottom quarks, \ttbar searches will be more sensitive than dijet searches for models with preferential couplings to \ttbar.

\subsection{Spin-2 colour singlet}
Spin-2, colour-singlet bosons are predicted in models that postulate extra dimensions of space leading to Kaluza–Klein (KK) excitations of the graviton. This search considers a Randall-Sundrum (RS) model with a single warped extra dimension, where the SM fields are in the bulk and the fermions are localised appropriately to explain the flavour structure of the SM~\cite{Randall:1999ee,Agashe:2007zd,Fitzpatrick:2007qr}. The type of KK graviton predicted in this RS model is referred to as a ``bulk'' RS graviton, or \grav, and it is characterised by its mass, \(m_{G_{KK}}\), and the dimensionless coupling of the KK gravitons to the SM fields, \(k/\overline{M}_{\mathrm{Pl}}\), where \(\overline{M}_{\mathrm{Pl}}=M_{\mathrm{Pl}}/8\pi\) is the reduced Planck mass and \(k\) is the curvature of the warped extra dimension. Larger values of \(k/\overline{M}_{\mathrm{Pl}}\) yield a larger production cross-section and a larger resonance width. In RS models with SM fields in the bulk, values of \(k/\overline{M}_{\mathrm{Pl}} \sim 1\) ensure that higher-order corrections to this model can be neglected~\cite{Agashe:2007zd}. For these KK gravitons, decays into light fermions are suppressed and the branching ratio to photons is negligible. In this and previous ATLAS papers, \(k/\overline{M}_{\mathrm{Pl}}= 1.0\), a choice that maximises the production cross-section while keeping the resonance width below the experimental resolution. In the mass range 0.4--3~\TeV considered for the KK graviton in this paper, the width varies between 3\% and 6\% of the mass. The primary production mode is \(gg\) fusion, while the primary decay modes are those to \ttbar, \(WW\), \(ZZ\), and \(hh\). The branching ratio of the \grav decay into a \ttbar pair increases rapidly from 18\% to 50\% for masses between 0.4 and 0.6~\TeV, reaching a plateau at 68\% for masses larger than 1~\TeV.

A previous search in the 1-lepton channel using 36~\ifb of ATLAS data~\cite{EXOT-2015-04} excluded bulk RS gravitons with masses in the range of 0.45--0.65~\TeV. A complementary search on the same ATLAS data sample targeting the decays \(G_{KK} \rightarrow WW,ZZ\)~\cite{EXOT-2017-31} excluded masses up to 2.3~\TeV. A dijet search~\cite{EXOT-2019-03} considered the \grav benchmark with \(k/\overline{M}_{\mathrm{Pl}}=0.2\) decaying into a bottom quark--anti-quark pair, and excluded these signals with masses up to 2.8~\TeV. A CMS search on 35.6~\ifb of \sqrtS data targeting the decay \(G_{KK} \rightarrow ZZ\)~\cite{CMS-B2G-16-023} excluded the \grav signal with \(k/\overline{M}_{\mathrm{Pl}}=0.5\) up to 0.8~\TeV\footnote{The \grav benchmark was not considered by the CMS Collaboration in their latest \ttbar resonance searches~\cite{CMS-B2G-17-017}.}. The model used in this paper and the papers mentioned above is itself ruled out by precision electroweak tests; nevertheless, it provides a useful generic benchmark that allows interpretations assuming some custodial symmetry to circumvent the electroweak constraints.

\subsection{Spin-1 colour octet}
Spin-1 colour-octet bosons are predicted in models with extra dimensions of space leading to KK excitations of the gluon~\cite{Lillie:2007yh,Ask:2011zs}. The benchmark model chosen in this paper, which is consistent with the benchmark choices in previous ATLAS publications, is an RS model with a single warped extra dimension, where the SM fields propagate in the bulk. The parameters of the model are also consistent with that of the \grav model described above. The primary production mode of these KK gluons, referred to as \glu, is via \(q\bar{q}\) annihilation. The strong coupling of the \glu to light quarks is set to \(g_q=0.2 g_s\), where \(g_s\) is the SM gluon coupling to quarks and the couplings correspond to the configuration given in Eq.~(3) of Ref.~\cite{Lillie:2007yh}. The left-handed coupling to the top quark is fixed at \(g_{t,\mathrm{L}} = g_s\), and the right-handed coupling to the top quark, \(g_{t,\mathrm{R}}\), is chosen to obtain a width of 30\% to obtain benchmark predictions for a broad resonance. For the configuration described here, this sets \(g_{t,\mathrm{L}}=g_{t,\mathrm{R}}\). This is fully consistent with the previous ATLAS searches for a \glu signal~\cite{EXOT-2015-04, EXOT-2016-24}. The branching fraction to \ttbar in this case amounts to 92.5\%.

A previous search in the 1-lepton channel using 36~\ifb of ATLAS data~\cite{EXOT-2015-04} excluded KK gluon masses lower than 3.7~\TeV in this benchmark model. The CMS Collaboration searched for similar resonances in 35.9~\ifb of \sqrtS data~\cite{CMS-B2G-17-017} using a combination of the three \ttbar decay channels, excluding \glu signals with masses up to 4.55~\TeV. This is a slightly different benchmark model than that used in this paper, which leads to a natural width of 15--20\% and a smaller production cross-section times branching fraction~\cite{Agashe:2006hk}.


%
\section{ATLAS detector}
\label{sec:detector}
%

The ATLAS detector~\cite{PERF-2007-01} at the LHC covers nearly the entire solid angle around the collision point.\footnote{ATLAS uses a right-handed coordinate system with its origin at the nominal interaction point (IP) in the centre of the detector and the \(z\)-axis along the beam pipe. The \(x\)-axis points from the IP to the centre of the LHC ring, and the \(y\)-axis points upwards. Polar coordinates \((r,\phi)\) are used in the transverse plane, \(\phi\) being the azimuthal angle around the \(z\)-axis. The pseudorapidity is defined in terms of the polar angle \(\theta\) as \(\eta = -\ln \tan(\theta/2)\) and is equal to the rapidity $ y = \frac{1}{2} \ln \left( \frac{E + p_z}{E - p_z} \right) $ in the relativistic limit. Angular distance is measured in units of \(\Delta R \equiv \sqrt{(\Delta y)^{2} + (\Delta\phi)^{2}}\).} It consists of an inner tracking detector surrounded by a thin superconducting solenoid, electromagnetic and hadronic calorimeters, and a muon spectrometer incorporating three large superconducting air-core toroidal magnets.

The inner-detector system (ID) is immersed in a \qty{2}{\tesla} axial magnetic field  and provides charged-particle tracking in the range \(|\eta| < 2.5\). The high-granularity silicon pixel detector covers the vertex region and typically provides four measurements per track, the first hit generally being in the insertable B-layer (IBL) installed before Run~2~\cite{ATLAS-TDR-19,PIX-2018-001}. It is followed by the SemiConductor Tracker (SCT), which usually provides eight measurements per track. These silicon detectors are complemented by the transition radiation tracker (TRT), which enables radially extended track reconstruction up to \(|\eta| = 2.0\). The TRT also provides electron identification information
based on the fraction of hits (typically 30 in total) above a higher energy-deposit threshold corresponding to transition radiation.

The calorimeter system covers the pseudorapidity range \(|\eta| < 4.9\). Within the region \(|\eta|< 3.2\), electromagnetic calorimetry is provided by barrel and endcap high-granularity lead/liquid-argon (LAr) calorimeters, with an additional thin LAr presampler covering \(|\eta| < 1.8\) to correct for energy loss in material upstream of the calorimeters. Hadronic calorimetry is provided by the steel/scintillator-tile calorimeter, segmented into three barrel structures within \(|\eta| < 1.7\), and two copper/LAr hadronic endcap calorimeters. The solid angle coverage is completed with forward copper/LAr and tungsten/LAr calorimeter modules optimised for electromagnetic and hadronic energy measurements, respectively.

The muon spectrometer (MS) comprises separate trigger and high-precision tracking chambers measuring the deflection of muons in a magnetic field generated by the superconducting air-core toroidal magnets. The field integral of the toroids ranges between \num{2.0} and \qty{6.0}{\tesla\metre} across most of the detector. Three layers of precision chambers, each consisting of layers of monitored drift tubes, cover the region \(|\eta| < 2.7\), complemented by cathode-strip chambers in the forward region, where the background is highest. The muon trigger system covers the range \(|\eta| < 2.4\) with resistive-plate chambers in the barrel, and thin-gap chambers in the endcap regions.

The luminosity is measured mainly by the LUCID--2~\cite{LUCID2} detector that records Cherenkov light produced in the quartz windows of photomultipliers located close to the beampipe.

Events were selected by the first-level trigger system implemented in custom hardware, followed by selections made by algorithms implemented in software in the high-level trigger~\cite{TRIG-2019-04}. The first-level trigger accepted events from the \qty{40}{\MHz} bunch crossings at a rate close to \qty{100}{\kHz}, which the high-level trigger further reduced in order to record complete events to disk at about \qty{1.25}{\kHz}.

A software suite~\cite{SOFT-2010-01} is used in data simulation, in the reconstruction and analysis of real and simulated data, in detector operations, and in the trigger and data acquisition systems of the experiment.


%
\section{Data and simulated event samples}
\label{sec:datamc}
%

Data from \(pp\) collisions at \sqrtS corresponding to an integrated luminosity of 140~\ifb, collected in the years 2015 to 2018 with the ATLAS detector, are analysed in the context of this search. The uncertainty in the combined 2015--2018 integrated luminosity is 0.83\%~\cite{DAPR-2021-01}, obtained primarily using the LUCID-2 detector~\cite{LUCID2}, and complemented by measurements provided by the inner detector and the calorimeters. The average number of interactions per bunch crossing (pile-up) in this data sample is \averagemu\(=33.7\). All detector subsystems were required to be operational during data taking~\cite{DAPR-2018-01}.

Candidate events for both the 1- and 2-lepton channels are selected using single-muon~\cite{TRIG-2018-01} and single-electron~\cite{TRIG-2018-05} triggers. These triggers require a muon or an electron with momentum transverse to the beamlime, \pT, or energy transverse to the beamline, \et, respectively, above a certain threshold and satisfying certain quality and, for some triggers, lepton isolation requirements. The triggers with the lowest \pT (\et) threshold require \(\pT (\et) > 26\)~\GeV for muons (electrons) in 2016--2018 and \(\pT (\et) > 20 (24)\)~\GeV in 2015, and include a lepton isolation requirement that is not applied for triggers with higher thresholds. The trigger efficiency is mostly constant in the transverse momentum for leptons with \(\pT > 28\)~\GeV, but can vary with \(\eta\).

Monte Carlo (MC) simulated event samples are used to model the signal and background processes. Unless otherwise stated, the top quark mass was always set to \(m_{\mathrm{top}}=172.5\)~\GeV; samples that used \Pythia~\cite{Sjostrand:2007gs} for the generation used the A14 set of tuned parameters (tune)~\cite{ATL-PHYS-PUB-2014-021} with the \NNPDF[2.3lo]~\cite{Ball:2012cx} set of parton distribution functions (PDFs); and all samples that did not use \SHERPA~\cite{Bothmann:2019yzt} used \EVTGEN~\cite{Lange:2001uf} to model the decays of bottom and charm hadrons. The ATLAS simulation infrastructure~\cite{SOFT-2010-01} was used to simulate the detector and its response. Signal and nominal background samples as well as several alternative background samples used to assess systematic uncertainties were produced with a detailed \GEANT~\cite{Agostinelli:2002hh} detector simulation. A faster simulation based on a parameterisation of the calorimeter response and \GEANT for the other detector systems~\cite{SOFT-2010-01} was used for the remaining alternative background samples. Pile-up effects were modelled by overlaying minimum-bias events simulated using the soft QCD processes of \PYTHIA[8.186] with the \NNPDF[2.3lo] set of PDFs and the A3 tune~\cite{ATL-PHYS-PUB-2016-017}. The pile-up profiles match the ones of each data sample between 2015 and 2018. The same offline reconstruction methods used for data were applied to the simulated event samples. Corrections were applied to the simulated events to match the selection efficiencies, energy and mass scales and resolutions of reconstructed simulated particles to those measured in data control samples. An overview of the generator choices for the signal and background processes can be found in Table~\ref{tab:samples:generators}, with further details given in the following.
\begin{table}[t]
\centering
\caption{List of matrix element (ME) generators and the order of the strong coupling constant in the perturbative calculation, parton distribution function (PDF) sets, parton shower (PS) and hadronisation generator, and underlying event (UE) tune for the different signal and background processes.}
\label{tab:samples:generators}
\resizebox{\linewidth}{!}{\begin{tabular}{lccccc}
\toprule
Process 			& ME generator 		& ME order 		& PDF set (ME)	& PS and hadronisation 	& UE tune   \\
\midrule
Signal \zprimeTC		& \MADGRAPH[2.9.3]     	& LO 			& \NNPDF[2.3lo] & \PYTHIA[8.245]    	& A14       \\
Signal \glu		        & \MADGRAPH[2.9.3]     	& LO 			& \NNPDF[2.3lo] & \PYTHIA[8.245]    	& A14       \\
Signal \grav		        & \MADGRAPH[2.2.3]     	& LO 			& \NNPDF[2.3lo] & \PYTHIA[8.210]    	& A14       \\
\midrule
\ttbar			        & \POWHEGBOX[v2] 	& NLO 			& \NNPDF[3.0nlo] 	& \PYTHIA[8.230]	& A14 	    \\
Single top (\(Wt\))	        & \POWHEGBOX[v2] 	& NLO 			& \NNPDF[3.0nlo] 	& \PYTHIA[8.307]	& A14 	    \\
Single top (\textit{s}-channel)	        & \POWHEGBOX[v2] 	& NLO 			& \NNPDF[3.0nlo] 	& \PYTHIA[8.230]	& A14 	    \\
Single top (\textit{t}-channel)	        & \POWHEGBOX[v2] 	& NLO 			& \NNPDF[3.0nlo] 	& \PYTHIA[8.230]	& A14 	    \\
Diboson 			& \SHERPA[2.2.1/2.2.2]   & \MEPSatNLO	        & \NNPDF[3.0nnlo]    & \SHERPA		& author tune  \\
\Wjets			& \SHERPA[2.2.11]     	& \MEPSatNLO	        & \NNPDF[3.0nnlo]	& \SHERPA 		& author tune  \\
\Zjets			& \SHERPA[2.2.1/2.2.11] & \MEPSatNLO	        & \NNPDF[3.0nnlo]	& \SHERPA 		& author tune  \\
\ttbarV			& \MGNLO[2.3.3] 	& NLO 			& \NNPDF[3.0nlo]     & \PYTHIA[8.210]	& A14 	    \\
\ttbarH			& \POWHEGBOX[v2] 	& NLO 			& \NNPDF[3.0nlo]     & \PYTHIA[8.230]	& A14 	    \\
\bottomrule
\end{tabular}}
\end{table}

The dominant background process is SM \ttbar production. Smaller background components arise from single vector-boson (\textit{W}, \textit{Z}) production in association with hadronic jets, referred to as \(W/Z+\mathrm{jets}\), and single top-quark production (single top). There are also background processes in both channels that lead to at least one mis-identified or non-prompt lepton that satisfies the lepton identification and isolation criteria applied in this search, referred to as \textit{fakes} in this paper. These mostly originate from strong multijet production in the 1-lepton channel and from single-lepton \ttbar events in the 2-lepton channel. Single top and \Wjets production can also lead to these mis-identified or non-prompt leptons. Minor backgrounds from \ttbarV, \ttbarH and diboson (\(WW\), \(ZZ\), \(WZ\)) production are also considered. All background components, except the fakes in the 1-lepton channel (see Section~\ref{subsec:multijet_1l}), are modelled using MC simulation. Data-driven corrections are applied to the MC-based estimates of the \Wjets background in the 1-lepton channel (see Section~\ref{subsec:wjets_1l}) and the \Zjets background in the 2-lepton channel (see Section~\ref{subsec:zjets_2l}) to correct the \Wjets normalisation and \Zjets shape, respectively.

SM \ttbar production was generated at NLO accuracy in QCD using \POWHEGBOX[v2]~\cite{Nason:2004rx,Frixione:2007nw,Frixione:2007vw,Alioli:2010xd,Campbell:2014kua} with the \NNPDF[3.0nlo]~\cite{Ball:2014uwa} PDF set and the \hdamp parameter set to 1.5\,\mtop~\cite{ATL-PHYS-PUB-2016-020}.\footnote{The \hdamp parameter controls the \pT of the first additional emission beyond the leading-order Feynman diagram in the parton shower and therefore regulates the high-\pT emission against which the \ttbar system recoils.} The functional form of the renormalisation and factorisation scales was set to the default scale \(\sqrt{m_{\textrm{top}}^2 + p_{\textrm T}^2}\). The top quarks were decayed in \POWHEGBOX, thus preserving their spin correlations. The matrix element (ME) generator is interfaced with \PYTHIA[8.230]~\cite{Sjostrand:2014zea} to model the parton shower (PS), hadronisation, and the underlying event (UE).

Alternative SM \ttbar samples obtained with different generator choices and settings are used to estimate systematic uncertainties related to the modelling of this main background component. The details can be found in Section~\ref{subsec:unc_mod}.

The top-quark kinematics in the nominal and all alternative SM \ttbar samples were corrected to more accurate differential predictions calculated at next-to-next-to-leading-order (NNLO) QCD and NLO electroweak (EW) accuracy for a top-quark mass value of \mbox{\mtop=173.3~\GeV~\cite{Czakon:2017wor}}. The difference between the \mtop value used in the reweighting and the value of 172.5~\GeV used in the NLO MC generation was found to have no impact on the analysis sensitivity. The corrections were applied as event-by-event weights that were obtained via an iterative recursive reweighting procedure in the differential distributions of the (anti)top-quark \pT and \mttbar, referred to as \textit{NNLO reweighting}, following the same procedure as described in Ref.~\cite{EXOT-2020-25}. The systematic uncertainties related to the NNLO reweighting procedure are listed in Section~\ref{subsec:unc_mod}.

All reweighted \ttbar samples were additionally normalised to the cross-section prediction at NNLO in QCD, including the resummation of next-to-next-to-leading logarithmic (NNLL) soft-gluon terms calculated using \TOPpp[2.0]~\cite{Beneke:2011mq,Cacciari:2011hy,Baernreuther:2012ws,Czakon:2012zr,Czakon:2012pz,Czakon:2013goa,Czakon:2011xx}, corresponding to \(\mathrm{832\pm51~pb}\).

Single top production in the \(Wt\)-channel~\cite{Re:2010bp}, which constitutes the main single top contribution, was generated with \POWHEGBOX[v2] using the \NNPDF[3.0nlo] PDF set. The overlap between \ttbar and \(Wt\) production, arising due to NLO EW corrections to \(Wt\), was treated within the diagram removal (DR) scheme~\cite{Frixione:2008yi} and the renormalisation and factorisation scales were set to \(H_{\mathrm{T}}/2\), where \(H_{\mathrm{T}}\) denotes the scalar sum of the transverse momenta of all final-state particles in the event. The ME generator was interfaced with \PYTHIA[8.307]~\cite{Bierlich:2022pfr} for the PS, hadronisation, and the UE modelling. The sample was normalised to the theory cross-sections for inclusive \(Wt\), calculated at NLO accuracy in QCD with NNLL soft-gluon corrections and corresponding to \(71.7 \pm 3.8\,\unit{\pb}\)~\cite{Kidonakis:2010ux}.

Single top production in the \textit{s}-channel~\cite{Alioli:2009je} was generated with \POWHEGBOX[v2] at NLO in QCD using the five-flavour scheme and the \NNPDF[3.0nlo] PDF set. Single top production in the \textit{t}-channel was generated with \POWHEGBOX[v2]~\cite{Frederix:2012dh}, using the four-flavour scheme for the NLO ME calculations together with the corresponding PDF \NNPDF[3.0] set. For this process, the top-quark decays were simulated using \MADSPIN~\cite{Artoisenet:2012st}, preserving all spin correlations. For both processes, the ME generator was also interfaced with \PYTHIA[8.230]. The inclusive cross-sections were corrected to the theory predictions at NLO in QCD with
\HATHOR[2.1]~\cite{Aliev:2010zk,Kant:2014oha}. For the \textit{s}-channel process, this cross-section corresponds to \(6.35^{+0.23}_{-0.20}\,\unit{\pb}\) (\(3.97^{+0.19}_{-0.17}\,\unit{\pb}\)) for single top (single anti-top) production~\cite{Kidonakis:2010tc}, and for the \textit{t}-channel process, this cross-section corresponds to \(136.0^{+5.4}_{-4.6}\)\,pb (\(81.0^{+4.1}_{-3.6}\)\,pb)~\cite{Kidonakis:2011wy} for single top (single anti-top) production.

The background from \ttbarV (\(V=W,Z\)) production was generated at NLO accuracy in QCD with \MGNLO[2.3.3]~\cite{Alwall:2014hca} with the \NNPDF[3.0nlo] PDF set, interfaced with \PYTHIA[8.210] for the PS, hadronisation, and UE. The samples were normalised to the theory cross-section calculated at NLO in QCD~\cite{deFlorian:2016spz}. The background from \ttbarH production was generated at NLO accuracy in QCD with \POWHEGBOX[v2] with the \NNPDF[3.0nlo] PDF set, interfaced with \PYTHIA[8.230] for the PS, hadronisation, and UE. The samples were normalised to the theory cross-section calculated at NLO QCD and NLO EW accuracies~\cite{deFlorian:2016spz}.

The production of a single \textit{W} boson in association with hadronic jets (\Wjets) was simulated with \SHERPA[2.2.11]. The background from \Zjets production was simulated with \SHERPA[2.2.11] in the 1-lepton channel and with \SHERPA[2.2.1] in the 2-lepton channel. NLO ME calculations were used for up to two partons for all samples, and leading-order (LO) ME calculations for up to five (four) partons for the \SHERPA[2.2.11] (\SHERPA[2.2.1]) samples. The MEs were calculated with the \AtlasMC{Comix}~\cite{Gleisberg:2008fv} and \OPENLOOPS~\cite{Buccioni:2019sur,Cascioli:2011va,Denner:2016kdg} libraries. They were matched with the \SHERPA PS~\cite{Schumann:2007mg} using the \MEPSatNLO prescription~\cite{Hoeche:2011fd,Hoeche:2012yf,Catani:2001cc,Hoeche:2009rj} and the set of tuned parameters developed by the \SHERPA authors. The \NNPDF[3.0nnlo] set of PDFs was used for all \Vjets samples. The \Wjets and \Zjets samples were normalised to the theory cross-sections calculated at NLO accuracy in QCD~\cite{Catani:2009sm}. An additional data-driven normalisation correction was derived for the \Wjets background in the 1-lepton channel. Details can be found in Section~\ref{subsec:wjets_1l}.

Diboson events with fully leptonic and semileptonic decays were simulated using \SHERPA[2.2.1] and \SHERPA[2.2.2]~\cite{ATL-PHYS-PUB-2017-005}, respectively. The simulation includes off-shell effects and Higgs boson contributions, where appropriate. Events were simulated using ME calculations at NLO accuracy in QCD for up to one additional parton and at LO accuracy for up to three additional parton emissions. Samples for the loop-induced processes \(gg\rightarrow VV\) were simulated using LO-accurate ME calculations for up to one additional parton emission. The ME calculations were matched and merged with the \SHERPA PS based on the Catani--Seymour dipole factorisation~\cite{Gleisberg:2008fv,Schumann:2007mg} using the MEPS@NLO prescription. The virtual QCD corrections were provided by the \OPENLOOPS library. The \NNPDF[3.0nnlo] set of PDFs was used, along with the dedicated set of tuned PS parameters developed by the \SHERPA authors. The cross-sections from the generator were used for sample normalisation.

The fakes background in the 2-lepton channel is estimated using the single-lepton \ttbar MC sample, the single top MC samples and the \Wjets MC sample that contains two leptons. After requiring two reconstructed leptons, if at least one of the leptons is not matched to a prompt lepton at generator level, the event is considered to be part of the fakes background.

The production of each of the three signals decaying into \ttbar was modelled at LO in QCD with the \NNPDF[2.3lo] PDF set using \MADGRAPH[2.9.3] for the \zprimeTC and \glu signals, and \MADGRAPH[2.2.3] for the \grav signal. \PYTHIA[8.245] was used to model the PS and hadronisation of the \zprimeTC and \glu signals, and \PYTHIA[8.210] was used for the \grav signal. For the \zprimeTC signals, the generated values of the cross-section were scaled to the cross-section predictions obtained at NLO in QCD for the 1.2\% width~\cite{Bonciani:2015hgv}, whilst the cross-sections obtained with \MADGRAPH were used directly for the \glu and \grav samples.


%
\section{Object reconstruction}
\label{sec:objects}
%

Common event-quality criteria and object definitions are applied for both analysis channels, including standard data-quality requirements to select data events with the detector in good operating condition~\cite{DAPR-2018-01}. In addition, in each analysis channel, dedicated event selection criteria, which are specific to the objects and kinematics of interest in those final states, are applied as described in Section~\ref{sec:selection}.

Events are required to have at least one reconstructed \(pp\) interaction vertex with a minimum of two associated tracks with \(p_{\mathrm{T}}>0.5\)~\GeV. The primary vertex matched to the hard-scatter of the interaction is defined as the vertex with the highest sum of squared transverse momenta of associated tracks~\cite{ATL-PHYS-PUB-2015-026}. A set of baseline quality criteria are applied to reject events with non-collision backgrounds or detector noise~\cite{ATLAS-CONF-2015-029}.

Two levels of object identification requirements are defined for charged leptons and jets: baseline and signal. Baseline leptons and jets are selected with looser identification criteria, and are used in computing the missing transverse momentum and in resolving possible reconstruction ambiguities. Signal leptons and jets are a subset of the baseline objects with tighter quality requirements, which are used to define the search regions. Isolation criteria are used to discriminate between prompt leptons and leptons arising from semileptonic heavy-flavour decays or jets mis-identified as leptons.

Jets are reconstructed from particle-flow objects~\cite{PERF-2015-09} using the \antikt algorithm~\cite{Cacciari:2008gp,Fastjet} with a radius parameter \(R=0.4\). The particle-flow algorithm combines information about ID tracks and energy deposits in the calorimeters to form the input for jet reconstruction. A pile-up subtraction procedure~\cite{JETM-2018-05} is applied along with a global sequential calibration to account for flavour dependencies. To suppress jets arising from pile-up, a jet-vertex-tagging (JVT) technique using a multivariate likelihood~\cite{PERF-2014-03} is applied to jets with \(p_{\mathrm{T}}<60\)~\GeV and \(|\eta|<2.4\) to reject jets that do not orginate from the hard-scatter primary vertex. Baseline jets are selected by requiring \(p_{\mathrm{T}}>20\)~\GeV and \(|\eta|<4.5\). Signal jets are selected via the tighter requirements \(p_{\mathrm{T}}>30\)~\GeV and \(|\eta|<2.5\).

Jets with a larger radius parameter, \(R=1.0\), referred to as \largeR jets in the following, are used as proxies for hadronically-decaying top quarks with a large Lorentz boost. They are reconstructed from three-dimensional topological clusters of energy deposits in the calorimeter using the \antikt algorithm. The topological clusters are calibrated to the hadronic energy scale with the local cluster weighting (LCW)~\cite{PERF-2014-07} procedure to account for the non-compensating calorimeter response and the energy loss in dead material or due to out-of-cluster leakage of deposited energy. The \largeR jets are trimmed~\cite{Krohn:2009th} to reduce contributions from pile-up and soft interactions by reclustering the jet constituents into subjets, using the \(k_t\) algorithm~\cite{Catani:1991hj,Ellis:1993tq,Catani:1993hr} with a radius parameter \(R=0.2\), and discarding constituents belonging to subjets with \pT less than 5\% of the \pT of the parent jet~\cite{JETM-2018-06}. The four-momenta of \largeR jets are reconstructed from the momentum vectors of the remaining subjets and calibrated using simulation~\cite{JETM-2018-02}. The \largeR jets used are required to have \(p_{\mathrm{T}}>350\)~\GeV and \(|\eta| < 2.0\).

\LargeR jets originating from a boosted hadronically-decaying top quark are identified using a multivariate classification algorithm based on a deep neural network (DNN)~\cite{JETM-2018-03} that takes 13 jet moments, including the jet mass and \pT, as inputs for training. A DNN score between zero and one is obtained, with top-quark-initiated jets having values close to one and light-quark- and gluon-initiated jets having values close to zero. The top-tagging requirements used are optimised for ``inclusive'' jets containing some but not necessarily all of the decay products of the hadronically-decaying top quark. A \largeR jet is considered ``top-tagged'' using a requirement on the DNN output score that corresponds to an identification efficiency of 80\% for inclusive top-jets, evaluated in simulated SM \ttbar events. The rejection factor for light-flavour quark and gluon jets varies as a function of the \largeR jet \pT, from approximately 60 at 350~\GeV to approximately 10 at 3000~\GeV, where the rejection factor is defined as the reciprocal of the probability of mistakenly identifying a light-quark-initiated jet or a gluon-initiated jet as a top-quark-initiated jet. Scale factors are applied to all jets satisfying the top-tagging requirement in simulated events to correct for possible efficiency differences between simulated event samples and data~\cite{JETM-2018-03}.

\SmallR jets in the range \(|\eta|<2.5\) are identified as likely to contain a \textit{b}-hadron, henceforth called \textit{b}-tagged, using the DL1r algorithm~\cite{FTAG-2019-07}. This algorithm is based on a multivariate classification technique with a DNN combining information from the impact parameters of tracks and topological properties of secondary and tertiary decay vertices reconstructed from the tracks associated with the jet. The \textit{b}-tagged jets are selected using an operating point corresponding to an efficiency of 77\% for identifying true \textit{b}-jets in simulated SM \ttbar events. This operating point corresponds to a rejection factor of 6 for $c$-jets and of 134 for light-flavour jets. Correction factors are applied to the simulated event samples to compensate for differences between data and simulation in the \textit{b}-tagging efficiency for \textit{b}-, \textit{c}-, and light-flavour jets~\cite{FTAG-2018-01,FTAG-2020-08,FTAG-2019-02}. The correction for \textit{b}-jets is derived from \ttbar events with final states containing two leptons, and the corrections are consistent with unity within uncertainties at the level of a few percent over most of the jet \pT range. The corrections for \textit{c}-jets and light-flavour jets are derived in semileptonic \ttbar events and \Zjets events, respectively.

Muon candidates are reconstructed from matching tracks in the ID and MS, refined through a global fit that uses the hits from both subdetectors~\cite{MUON-2018-03}. Baseline muons must have \(p_{\mathrm{T}}>10\)~\GeV and \(|\eta|<2.5\), and satisfy a set of medium identification criteria. Additionally, the longitudinal impact parameter is required to satisfy \(|z_0\sin\theta|<0.5\)~mm. Signal muons are required to have \(p_{\mathrm{T}}>25\)~\GeV and additionally satisfy the following requirement on the transverse impact parameter \(d_0\) and its uncertainty \(\sigma_{d_0}\): \(|d_0/\sigma_{d_0}|<3\). They are also required to be isolated using the requirement that the sum of the transverse momenta of the tracks within a variable-radius cone around the muon direction, excluding the muon track, be less than 6\% of the transverse momentum of the muon~\cite{MUON-2018-03}. The track isolation cone size is given by the minimum of \(R = 10\;\mathrm{GeV}/p_{\mathrm{T}}^{\mu}\) and \(R=0.3\), where \(p_{\mathrm{T}}^{\mu}\) is the muon \pt~\cite{TOPQ-2012-18}.

Electron candidates are reconstructed from energy deposits in the electromagnetic calorimeter matched to a charged-particle track in the ID~\cite{EGAM-2018-01}. The track is required to be matched to the primary vertex by imposing the requirement \(|z_0\sin\theta|<0.5\)~mm. Electron candidates are required to be within \(|\eta|<2.47\), excluding the transition region between the barrel and endcap calorimeters (\(1.37<|\eta|<1.52\)). Baseline electrons are required to satisfy \(p_{\mathrm{T}}>10\)~\GeV and fulfil loose identification criteria, using a likelihood-based discriminant that combines information about tracks in the ID and energy deposits in the calorimeter system~\cite{EGAM-2018-01}. The number of hits in the innermost pixel layer is used to discriminate between electrons and converted photons. Signal electrons are required to also satisfy \(p_{\mathrm{T}}>25\)~\GeV and the tight likelihood identification criteria~\cite{EGAM-2018-01}. Additionally, signal electrons must fulfil \(|d_0/\sigma(d_0)|<5\). The same variable-cone isolation requirement as with muons is imposed on signal electrons, with the exception that the maximum cone radius is set to 0.2.

An overlap removal procedure is applied to resolve the reconstruction ambiguities between electrons, muons, and \smallR jets. First, if an electron shares the same ID track with another electron, the electron with the lower transverse momentum is discarded. Electron candidates sharing the same inner-detector track as a muon are rejected, as they are likely mis-identified energy deposits from bremsstrahlung photons emitted by the muon. Next, the closest jet to an electron is rejected if it lies within \(\Delta R=0.2\) of the electron. Similarly, jets are rejected if they are within \(\Delta R=0.2\) of a muon if the jet has fewer than three associated tracks or the muon is matched to the jet through ghost association~\cite{Cacciari:2008gn}. Next, to reduce the background contributions due to muons from heavy-flavour decays inside jets, muons are discarded if they are separated from the nearest jet by \(\Delta R < 0.04 + 10\;\mathrm{GeV}/p_{\mathrm{T}}^{\mu}\). For electrons, an electron-in-jet subtraction method based on that employed in Ref.~\cite{TOPQ-2012-18} is used to suppress backgrounds from leptonic heavy-flavour decays inside jets, while maintaining a high reconstruction efficiency for electrons from highly collimated leptonic decays of boosted top quarks, which typically result in an electron close to a \textit{b}-jet. If an electron is found within \(\Delta R=0.4\) of a jet, the four-momentum of the electron is subtracted from that of the jet, and the \pT of the electron-subtracted jet is calculated. A \pT threshold is defined using simulated \(Z\rightarrow ee\) events and if the electron-subtracted jet has a \pT below this threshold, the jet is discarded and the electron is retained. If the electron-subtracted jet has a \pT above this threshold, the electron is discarded and the original jet is retained if the \(\Delta R\) between the electron and the electron-subtracted jet is less than 0.2. Both the electron and the electron-subtracted jet are retained if the \(\Delta R\) between them is greater than 0.2.

The missing transverse momentum \ptmiss, with magnitude \met, is calculated as the negative vectorial sum of the transverse momenta of all baseline reconstructed objects (electrons, muons, \smallR jets and photons~\cite{EGAM-2018-01}) and a soft term. The soft term includes all tracks associated with the primary vertex but not matched to any reconstructed physics object. Tracks not associated with the primary vertex are not considered in the \met calculation, improving the \met resolution by suppressing the effect of pile-up~\cite{JETM-2020-03}.


%
\section{Event selection and categorisation}
\label{sec:selection}
%

Events satisfying the single-electron and single-muon trigger requirements outlined in Section~\ref{sec:datamc}, and the data-quality requirements described in Section~\ref{sec:objects}, are classified as belonging to the 1- or 2-lepton channels based on the number of signal electrons or muons they contain. The lepton requirements applied in this context are designed to ensure orthogonality between the two channels. Events in the 1-lepton channel are further classified into orthogonal event categories based on whether their detector signature is compatible with the presence of either a merged or resolved hadronic top-quark decay, and based on the number of associated \textit{b}-jets. Angular requirements are used to define orthogonal signal and validation regions in both the 1- and 2-lepton channels. The latter have a similar background composition to the corresponding signal regions but are depleted in the expected signal processes. They serve to validate the modelling of the dominant background from SM \ttbar production.

\subsection{Event selection and categorisation in the 1-lepton channel}
\label{subsec:selection_1l}
Events are required to contain exactly one signal electron or muon with a minimum \(p_{\mathrm{T}}^{\ell}>28\)~\GeV. This threshold is chosen to ensure that events are selected from the trigger efficiency plateau. The electron or muon is also required to be matched, within \(\Delta R<0.15\), to the respective trigger object. Events with a second signal electron or muon (with \(p_{\mathrm{T}}>25\)~\GeV, see Section~\ref{sec:objects}) are vetoed. Regions are not split according to the flavour of the selected lepton because it was found that such a splitting would not increase the sensitivity of the analysis.

The detector signature of a \ttbar decay in the 1-lepton channel involves the presence of a substantial amount of missing transverse momentum from the leptonically-decaying \textit{W} boson. To suppress background from the fakes that can cause small amounts of \MET via leptonic decays in heavy-flavour jets or mis-measured jet energies, the requirement \(E_{\mathrm{T}}^{\textrm{miss}}>20\)~\GeV is imposed. Additionally, events are required to fulfil \(E_{\mathrm{T}}^{\textrm{miss}}+m_{\mathrm{T}}^{W}>60\)~\GeV, where the transverse mass of the selected lepton and the \ptmiss, referred to as the \textit{W} transverse mass, or \mwt, is defined as:
\begin{equation}
\mwt=\sqrt{2p_{\mathrm{T}}^{\ell}E_{\mathrm{T}}^{\textrm{miss}}(1-\cos\Delta\phi(p_{\mathrm{T}}^{\ell}, \vec{p}_{\mathrm{T}}^{\textrm{miss}}))}\,. \nonumber
\end{equation}
All events must contain at least one \textit{b}-tagged \smallR jet.

Events are further categorised based on their detector signature as containing either a merged hadronic top-quark decay (``merged category''), reconstructed as a single \largeR jet, or a resolved hadronic top-quark decay (``resolved category''), resulting in three individual (\smallR) jets. To obtain orthogonal signal regions, only events that do not meet the requirements for the merged-topology selection are considered for the resolved-topology selection. The preference for the merged- over the resolved-topology selection is based on its superior \mtt resolution (see Section~\ref{subsec:reco_1l}).

Events in the merged category must contain at least one \smallR jet with a distance \(\Delta R<2.0\) to the selected lepton. If only one jet is found, it is identified as the \textit{b}-jet candidate from the leptonic top-quark decay. If more than one jet is found within \(\Delta R<2.0\) from the selected lepton, the \textit{b}-jet candidate is chosen as the jet with the highest \pT among all \textit{b}-tagged candidate jets. If none of the candidate jets is \textit{b}-tagged, the jet with the highest \pT is chosen instead. In addition, exactly one top-tagged \largeR jet must be present in the event with a distance \(\Delta R>1.5\) to the lepton and to the selected \textit{b}-jet candidate from the leptonic top-quark decay.

Only events failing the merged selection are considered for the resolved-topology selection. A fully resolved hadronic top-quark decay is expected to result in three \smallR jets, one of which is a \textit{b}-jet, in addition to a \textit{b}-jet from the leptonic top-quark decay. Hence, events with a resolved decay topology are required to contain at least four selected \smallR jets. The \ttbar system is then reconstructed using a \(\chi^2\) algorithm, defined in Section~\ref{subsubsec:reco_1l:resolved}. Only well-reconstructed events, defined via the requirement \(\textrm{log}_{10}(\chi^2)<0.9\), are kept for further analysis. This requirement has an efficiency of around 60\% for SM \ttbar events. All selection criteria are summarised in Table~\ref{tab:selection_1l}.
\newenvironment{DIFnomarkup}{}{}
\begin{DIFnomarkup}
\begin{table}[t]
\caption{Summary of the event selection criteria for the merged and resolved categories in the 1-lepton channel.}
\label{tab:selection_1l}
\centering
\begin{tabular}{l | l  }
\toprule
\textbf{Selection} & \textbf{Criteria}\\
\midrule
\multicolumn{2}{c}{\textbf{Common selection}} \\
\midrule
Event cleaning & All detector components with acceptable conditions \\
Single-lepton trigger & Single-electron or single-muon trigger \\
\multirow{2}{*}{Exactly one lepton} & Exactly one \(e\) or \(\mu\) with \(p_{\mathrm{T}}>28\)~\GeV \\
& Veto second \(e\) or \(\mu\) with \(p_{\mathrm{T}}>25\)~\GeV \\
\met & \(E_{\mathrm{T}}^{\mathrm{miss}}>20\)~\GeV \\
\met + \mwt & \(E_{\mathrm{T}}^{\mathrm{miss}}+m_{\mathrm{T}}^W>60\)~\GeV \\
\textit{b}-tagging & \(\geq1\) \textit{b}-tagged \smallR jet \\
\midrule
\multicolumn{2}{c}{\textbf{Merged topology selection}} \\
\midrule
\LargeR jet & \(\geq1\) \largeR jet with \(p_{\mathrm{T}}>350\)~\GeV \\
Candidate top-jet (hadronic decay) & \(=1\) \largeR jet passing top-tagging requirements \\
Candidate \textit{b}-jet (leptonic decay) & \(\geq1\) \smallR jet with \(\Delta R (\ell,\mathrm{jet})<2.0\) \\
\multirow{3}{*}{No overlap between objects}   & \(\Delta R(\)candidate b-jet,\,\(\ell)<2.0\) \\
& \(\Delta R(\)candidate top-jet, candidate b-jet\()>1.5\) \\
& \(\Delta R(\)candidate top-jet,\,\(\ell)>1.5\) \\
\midrule
\multicolumn{2}{c}{\textbf{Resolved topology selection}} \\
\midrule
Veto events passing merged selection & \\
\SmallR jets & \(\geq4\) \smallR jets with \(p_{\mathrm{T}}>25\)~\GeV \\
Well-reconstructed \ttbar system & \(\mathrm{log}_{10}(\chi^2)<0.9\) \\
\bottomrule
\end{tabular}
\end{table}
\end{DIFnomarkup}

In Figure~\ref{fig:accXeff}, the selection efficiency times acceptance for the \ljets final state is shown for the resolved, merged, and the combined resolved+merged categories, separately for the three benchmark scenarios described in Section~\ref{sec:models}. It is given as a function of \mttbeforeFSR, which is defined using the invariant mass of the top and antitop quarks at parton level, before final-state radiation (FSR), and allows access to the invariant mass of the partonic system immediately from the resonance decay. The product of the acceptance times efficiency is defined as the ratio of events satisfying the signal region requirements to the total number of generated events. For reference, the branching ratio for \ttbar to \ejets or \mujets final states is about 17\% for each lepton flavour, taking into account leptonic \(\tau\)-lepton decays~\cite{Workman:2022ynf}. The efficiency times acceptance is dominated by the merged topology selection for \ttbar invariant mass values \(m_{t\bar{t}}>1000\)~\GeV, while the resolved topology selection covers the region down to the kinematic threshold of the analysis at 400~\GeV. The selection efficiency times acceptance has a comparable \mttbar dependence for all three benchmark models with small differences arising due to differences between the spin of the intermediate resonance, which leads to differences in the angular distributions of the \ttbar decay products. The values vary from around 3\% for the lowest mass signals up to around 12\% for signals with masses between about 1 and 1.5~\TeV. They are consistent within a few \% between the different signal hypotheses.

\begin{figure}[t]
\centering
\subfloat[]{
\includegraphics[width=0.47\textwidth]{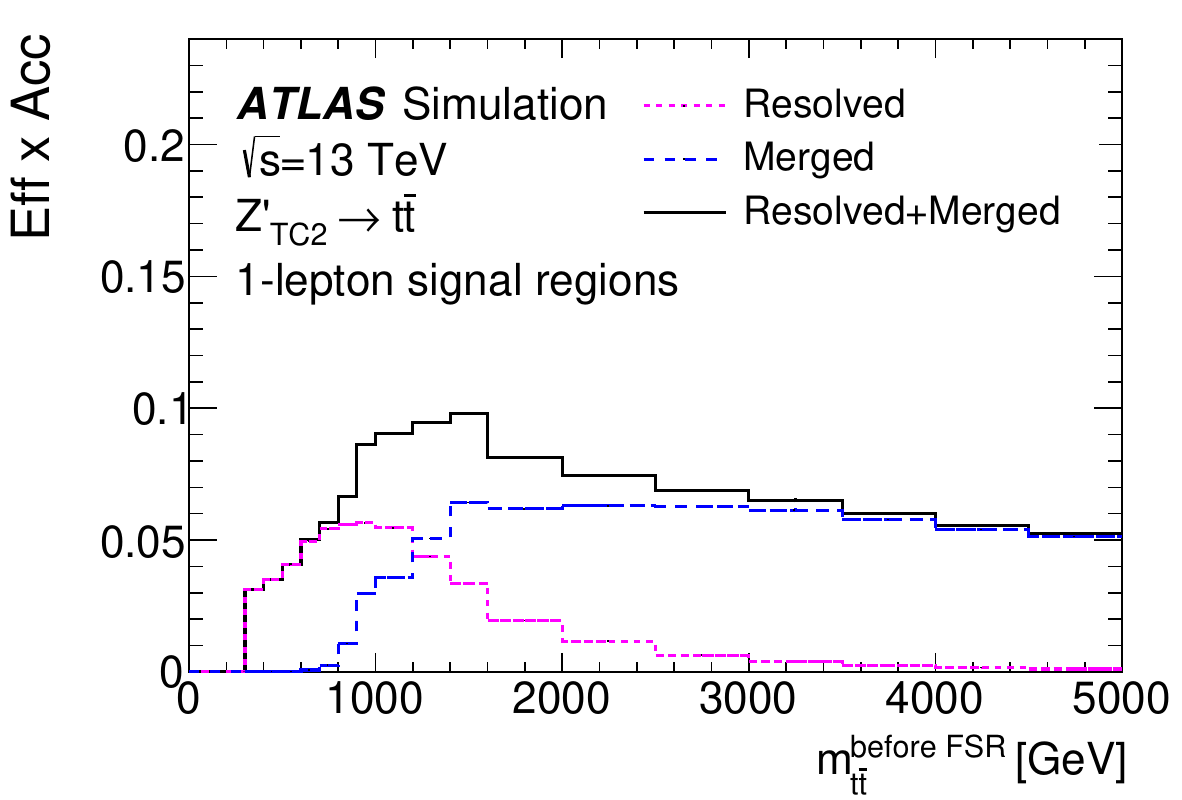}
\label{fig:accXeff_zprime}
}
\subfloat[]{
\includegraphics[width=0.47\textwidth]{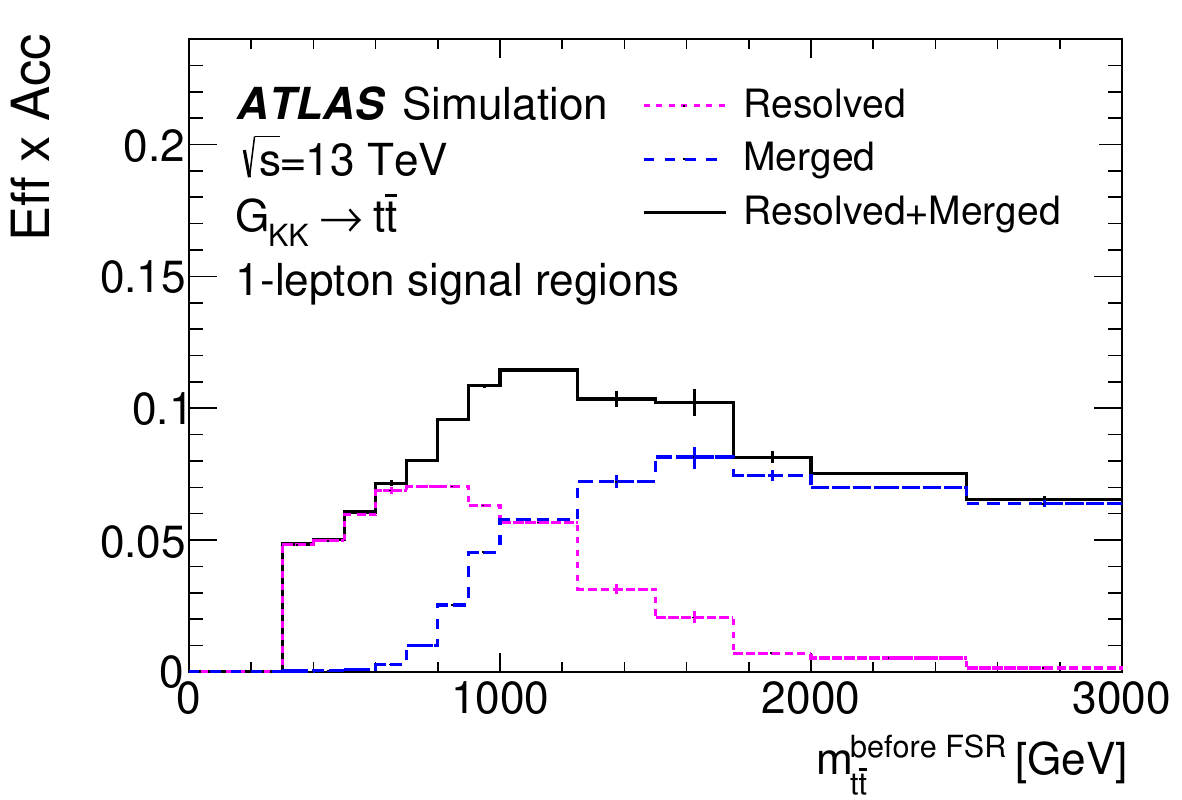}
\label{fig:accXeff_kkgrav}
}  \\
\subfloat[]{
\includegraphics[width=0.47\textwidth]{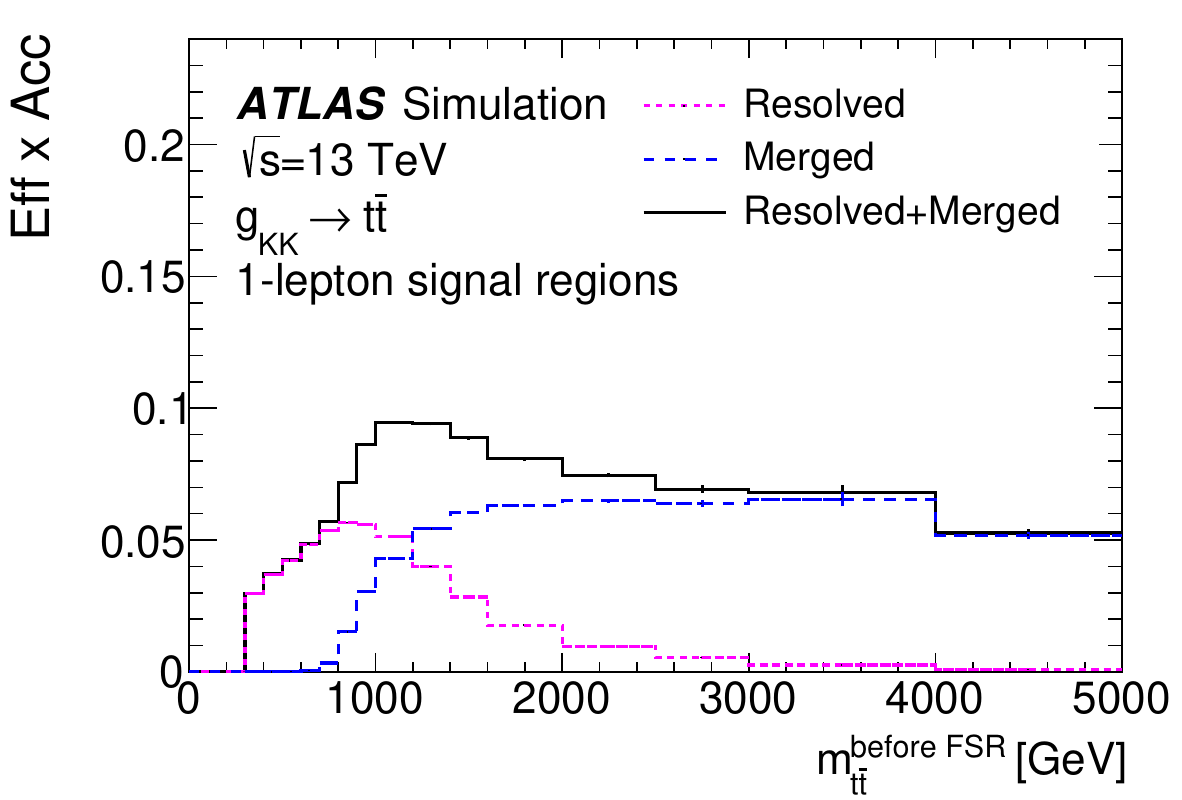}
\label{fig:accXeff_kkg}
}
\caption{Selection efficiency times acceptance (Eff x Acc) for the \ljets final state  as a function of the \ttbar invariant mass at the parton level before the emission of FSR, for the \protect\subref{fig:accXeff_zprime} \zprimeTC, \protect\subref{fig:accXeff_kkgrav} \grav, and \protect\subref{fig:accXeff_kkg} \glu signals. The selections in the resolved topology only (short-dashed magenta), the merged topology only (dashed blue) and the combined resolved+merged topology (solid black) are shown. The error bars correspond to the statistical uncertainty. All \ttbar decay modes are considered.}
\label{fig:accXeff}
\end{figure}
To further increase the search sensitivity, events passing the resolved topology selection are divided into separate sub-categories based on whether one or both of the reconstructed top quarks have an associated \textit{b}-tagged jet. The matching of \textit{b}-tagged jets to the leptonic or hadronic top quark is performed by checking whether the \smallR jet assigned as the \textit{b}-candidate jet to the leptonic side or one of the \smallR jets used for the reconstruction of the hadronic decay (based on the result of the \(\chi^2\) minimisation) is \textit{b}-tagged. The \textit{Resolved \(2b\)} category contains events in which both top-quark candidates have associated \textit{b}-tagged jets, while the \textit{Resolved \(1b\)} category contains events in which only the leptonically- or hadronically-decaying top-quark candidate has an associated \textit{b}-tagged jet. No \textit{b}-tagging categorisation is used for the merged topology as it provides minimal improvement in sensitivity.

The three 1-lepton categories (Merged, Resolved 2\textit{b}, and Resolved 1\textit{b}) are further sub-divided into orthogonal signal and validation regions based on the absolute value of the pseudorapidity difference between the selected lepton and the hadronic top quark, \dEtaTop. The pseudorapidity difference (between top-quark pairs or between particular decay products of each of them) is mostly sensitive to the different production mechanisms for SM \ttbar with respect to a \ttbar-resonance production process. Indeed, in SM \ttbar production, which proceed predominantly via gluon-initiated \textit{t}-channel diagrams, top quarks are preferentially produced along the incoming parton directions. On the other hand, for \ttbar resonances the primary diagrams are \textit{s}-channel ones (quark- or gluon-initiated, depending on the type of resonance), with a consequently more isotropic distribution. Spin-parity properties of the resonance can modify such pseudorapidity distributions, especially if the difference is taken between top-quark decay products such as charged leptons, but the separation power versus SM \ttbar stays mostly the same for all the considered signals. This makes \dEtaTop particularly useful to split between signal and validation regions, with validation regions having similar background composition as their corresponding signal regions, but being depleted in any of the signal processes considered. The separation power of these pseudorapidity variables grows with \mttbar, with less and less partonic energy left for boosting the \ttbar system along the \textit{z}-direction and consequently narrowing the pseudorapidity distribution of the \ttbar system.

These validation regions serve to validate the modelling of the dominant background from SM \ttbar production before the hypothesis testing in data. The \dEtaTop values chosen to define the signal and validation regions depend on the \mttbar range. This \mttbar-dependent \dEtaTop requirement is motivated by both statistical and kinematic considerations: it is relaxed at high \mttbar to retain sufficient expected event yields in the signal regions, while accounting for the increasing correlation between \dEtaTop and \mttbar, particularly for the SM \ttbar background, as discussed above. The values are summarised in Table~\ref{tab:deta_1l} and Figure~\ref{fig:dEta_1L} shows the distribution of \dEtaTop, inclusive in \mttbar. Very good agreement is observed between the data and the simulated backgounds already pre-fit.
\begin{table}[!ht]
\caption{Requirement on \dEtaTop used to define the signal regions for the resolved and merged decay topologies of the 1-lepton channel as a function of \mttbar. Values larger than those listed in the table correspond instead to the validation regions.}
\label{tab:deta_1l}
\begin{tabular}{cc}
\toprule
\mttbar range  & Resolved signal region \\
{[\(\)GeV]}        & requirement on \dEtaTop \\
\midrule
\(m_{t\bar{t}}<640\) 			& \(<1.5\) \\
\(640\leq m_{t\bar{t}}<800\)  	& \(<2.0\) \\
\(800\leq m_{t\bar{t}}<1280\)   & \(<2.5\) \\
\(1280\leq m_{t\bar{t}}<1700\)  & \(<3.0\) \\
\(m_{t\bar{t}}\geq1700\)	    & \(<3.5\) \\
\bottomrule
\end{tabular}
\begin{tabular}{cc}
\toprule
\mttbar  & Merged signal region \\
{[\GeV]}  & requirement on \dEtaTop \\
\midrule
\(m_{t\bar{t}}<1040\) 		   & \(<1.0\) \\
\(1040\leq m_{t\bar{t}}<1850\) & \(<1.5\) \\
\(1850\leq m_{t\bar{t}}<2500\) & \(<2.0\) \\
\(2500\leq m_{t\bar{t}}<3500\) & \(<2.5\) \\
\(m_{t\bar{t}}\geq3500\)	   & \(<3.0\) \\
\bottomrule
\end{tabular}
\end{table}
\begin{figure}[t]
\centering
\subfloat[]{
\includegraphics[width=0.4\textwidth]{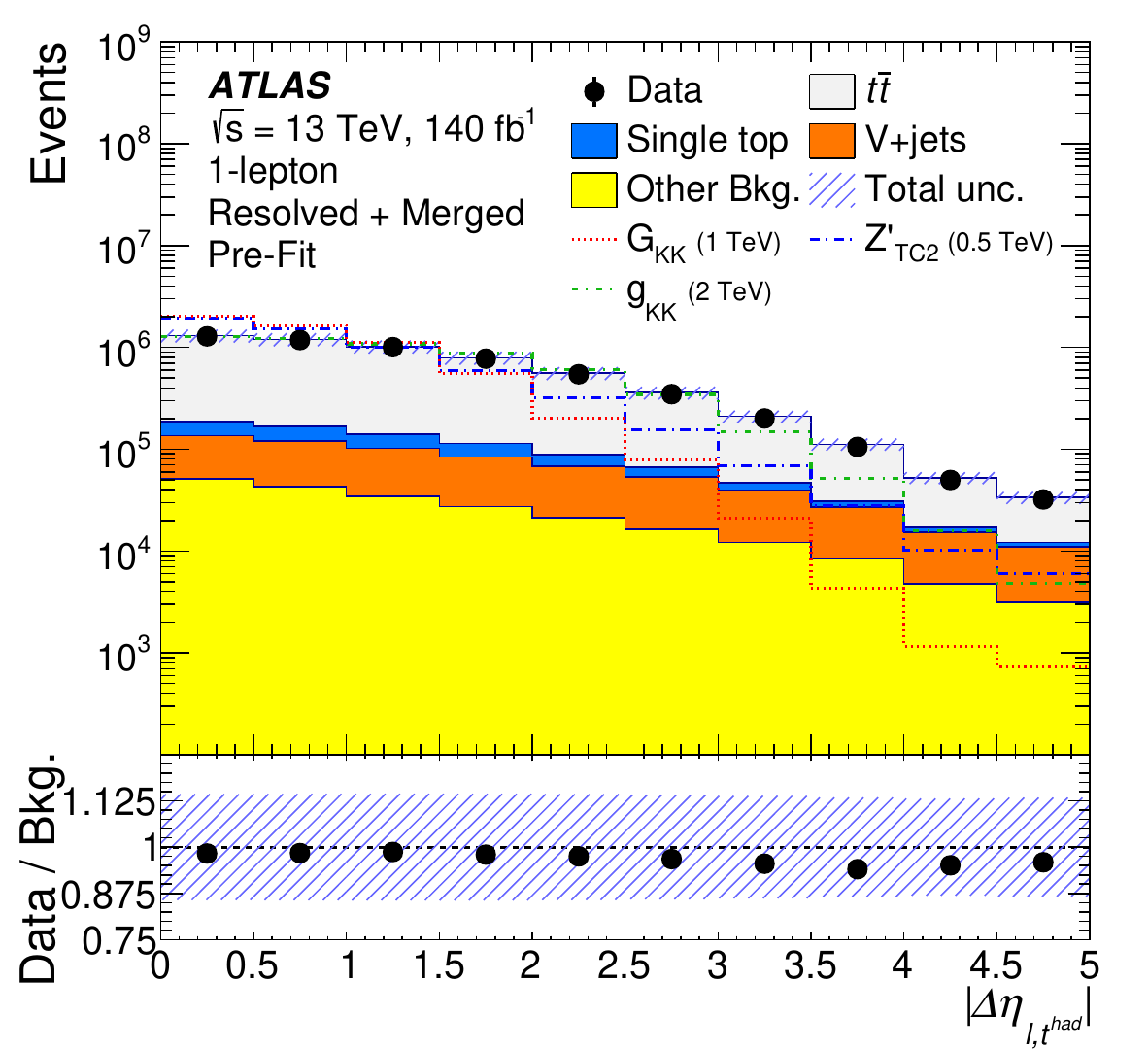}
\label{fig:dEta_1L}
}
\subfloat[]{
\includegraphics[width=0.4\textwidth]{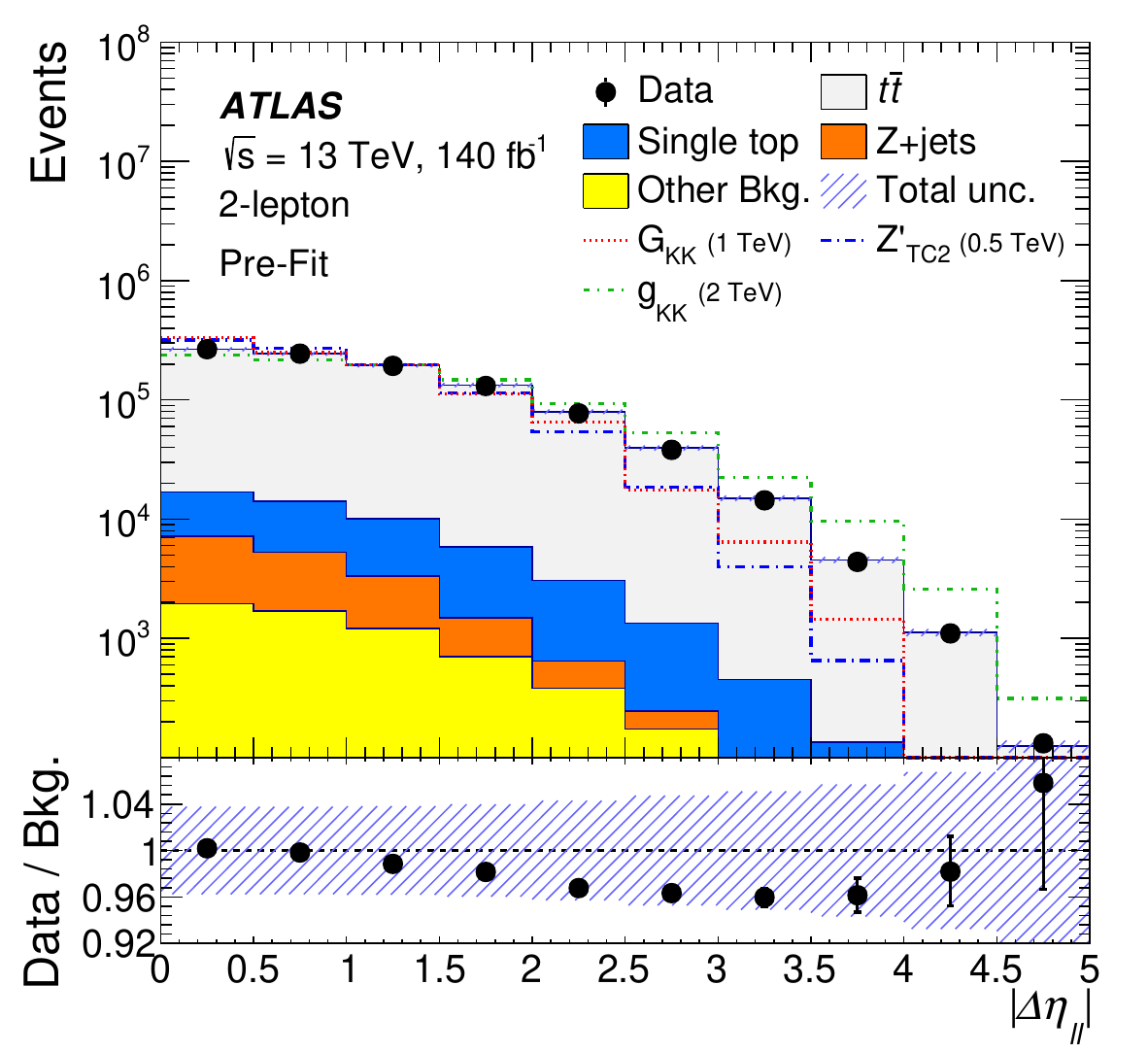}
\label{fig:dEta_2L}
} \\
\caption{Distribution of \protect\subref{fig:dEta_1L} the reconstructed \dEtaTop in the 1-lepton channel and \protect\subref{fig:dEta_2L} the reconstructed \dEta in the 2-lepton channel. The distributions and the uncertainty band are obtained before the final fit to the data (pre-fit). ``\Vjets'' refers to the \Zjets and \Wjets backgrounds, whilst ``Other Bkg.'' combines the \ttbarV, \ttbarH, diboson and fakes backgrounds together. A 0.5~\TeV \zprimeTC signal (blue dashed and dotted line), a 1~\TeV \grav signal (red dotted line) and a 2~\TeV \glu signal (green short-dashed and dotted line) are overlaid for illustration purposes, normalised to the total background. The lower panels show the ratio of the data to the total SM background prediction. The error bands include both the statistical and systematic uncertainties.}
\end{figure}

The highest SM \ttbar contribution (94.9\%) is found in the Resolved \(2b\) region due to the requirement of two \textit{b}-tagged jets. In the Resolved \(1b\) and Merged signal regions, the SM \ttbar production contributes to 79.2\% and 88.9\%, respectively, of the total SM background.

Additional control and validation regions enriched in hadronic jets falsely identified as leptons are defined for the data-driven estimate of the fakes background via the matrix method~\cite{EGAM-2019-01}. These are obtained by inverting the \met and/or \mwt requirements of the signal regions and additionally loosening the lepton identification and isolation requirements. Details are given in Section~\ref{subsec:multijet_1l}.

\subsection{Event selection and categorisation in the 2-lepton channel}
Candidate events for the 2-lepton channel are required to have exactly two signal leptons and at least two reconstructed jets. At least one of these leptons is required to have \(p_{\mathrm{T}}>28\)~\GeV and to match, within \(\Delta R<0.15\), the lepton of the same flavour reconstructed by the trigger algorithm. Additionally, at least one of the jets is required to be \textit{b}-tagged. Depending on the flavour of the charged leptons, three channels are defined: \(ee\), \(\mu\mu\) and \(e\mu\). Events satisfying these requirements are then further classified into signal and validation regions.

Events are required to contain two leptons with opposite-sign electric charge (OS). In the \(ee\) and \(\mu\mu\) channels, to further suppress the background from \Zyjets production, the dilepton invariant mass, \mll, is required to be greater than 15~\GeV and not within the range 81--101~\GeV (excluding the \textit{Z}-boson pole). The missing transverse momentum is also required to be larger than 45~\GeV. Additionally, events must meet the requirement that the invariant masses of the lepton--\textit{b}-jet pairs, \(m_{\ell b}\), be smaller than 150~\GeV for at least one of the two possible \textit{b}-jet to lepton assignments. In this context, the two \textit{b}-jets are either taken as the two \textit{b}-tagged jets with the largest \pT or, if only one jet in the event is \textit{b}-tagged, the \textit{b}-tagged jet and the jet with the highest \pT among the jets that are not \textit{b}-tagged. This requirement is meant to veto events where at least one of the lepton--\textit{b}-jet pairs does not originate from a top-quark decay. It effectively suppresses events from \(Wt\) production and \ttbar events in which at least one of the selected \textit{b}-jets does not originate from a top-quark decay but, for example, from initial-state radiation (ISR). It has a 64\% efficiency for SM \ttbar events with both top quarks decaying leptonically, and a 68\% efficiency for signal events with a resonance mass of 1000~\GeV. The event selection requirements are summarised in Table~\ref{tab:selection_2l} and the relative contribution of SM \ttbar production to the total SM background amounts to 94.7\%.
\begin{table}[t]
\caption{Summary of the event selection criteria for the signal regions in the 2-lepton channel.}
\label{tab:selection_2l}
\centering
\begin{tabular}{l | l  }
\toprule
\textbf{Selection} & \textbf{Criteria}\\
\midrule
Event cleaning & All detector components with acceptable conditions \\
Single-lepton trigger & Single-electron or single-muon trigger \\
\multirow{2}{*}{Exactly two leptons} & Exactly two (\(ee\), \(\mu\mu\), or \(e\mu\)) with \(p_{\mathrm{T}}>25\)~\GeV \\
& Leading lepton with \(p_{\mathrm{T}}>28\)~\GeV \\
\SmallR jets & \(\geq 2\) jets with \(p_{\mathrm{T}} > 25\)~\GeV \\
\textit{b}-tagging & \(\geq 1\) \textit{b}-tagged jet \\
Opposite-sign leptons & \(e^+e^-\), \(\mu^+\mu^-\), \(e^+\mu^-\), or \(e^-\mu^+\) \\
\met    & \(E_{\mathrm{T}}^{\mathrm{miss}}>45\)~\GeV (\(ee\) and \(\mu\mu\) channels only) \\
Dilepton invariant mass & \(15<m_{\ell\ell}<81\)~\GeV or \(m_{\ell\ell}>101\)~\GeV (\(ee\) and \(\mu\mu\) channels only) \\
Lepton-plus-\textit{b}-jet invariant mass & \(\mathrm{min}(m_{\ell b})<150\)~\GeV \\
\bottomrule
\end{tabular}
\end{table}

Five orthogonal regions based on the absolute value of the azimuthal angle between the two leptons, \dPhi, divided by \(\pi\) are defined: [0, 0.50], (0.50, 0.80], (0.80, 0.90], (0.90, 0.95], and (0.95, 1.00]. The \dPhi distribution is shaped by both the kinematics of the \ttbar system and the spin correlations of the top-quark pair. Increasing \mttbar leads to more boosted top quarks and hence larger values of \dPhi. On the other hand, at fixed \mttbar, spin-correlation effects in SM \ttbar production, together with the fact that charged leptons are emitted mostly along the spin directions of their parent top or anti-top quark in the respective rest frame, tend to favour configurations where charged leptons have smaller opening angles in the laboratory frame. Instead, heavy spin-1 or spin-2 resonances, due to the different possible spin configurations, produce on average charged-lepton pairs with larger opening angles. The exact \dPhiPi values used to define the five regions have been chosen to optimise the sensitivity of the 2-lepton channel to spin-1 resonances, while ensuring that each region contains enough data.

In analogy to the 1-lepton channel, events in these five regions are further divided into orthogonal signal and validation regions based on the absolute pseudorapidity difference between the two selected leptons, \dEta. The \dEta values chosen to define the signal and validation regions depend on the \mllbb range and are summarised in Table~\ref{tab:deta_2l}. Figure~\ref{fig:dEta_2L} shows the distribution of \dEta, inclusive in \mllbb. Very good agreement is observed between the data and the simulated backgrounds already pre-fit.
\begin{table}[!ht]
\caption{Requirement on \dEta used to define the signal regions as a function of \mllbb. Values larger than those listed in the table correspond instead to the validation regions.}
\label{tab:deta_2l}
\centering
\begin{tabular}{cc}
\toprule
\mllbb range & Signal region \\
{[\GeV]}      & requirement on \dEta  \\
\midrule
\(<450\)        & \(<1.0\) \\
450--1000     &  \(<1.5\) \\
1000--1500    & \(<2.0\) \\
1500--2500    & \(<2.5\) \\
\(>2500\)        & none \\
\bottomrule
\end{tabular}
\end{table}

The pre-fit distribution of \dPhiPi in the 2-lepton channel is shown in Figure~\ref{fig:dPhi:prefit:incl}, separately for the signal and validation regions. Each \dPhiPi bin corresponds to a signal or validation region, respectively. Excellent agreement is observed between the data and the SM yields already pre-fit, in both regions.
\begin{figure}[t]
\centering
\subfloat[]{
\includegraphics[width=0.4\textwidth]{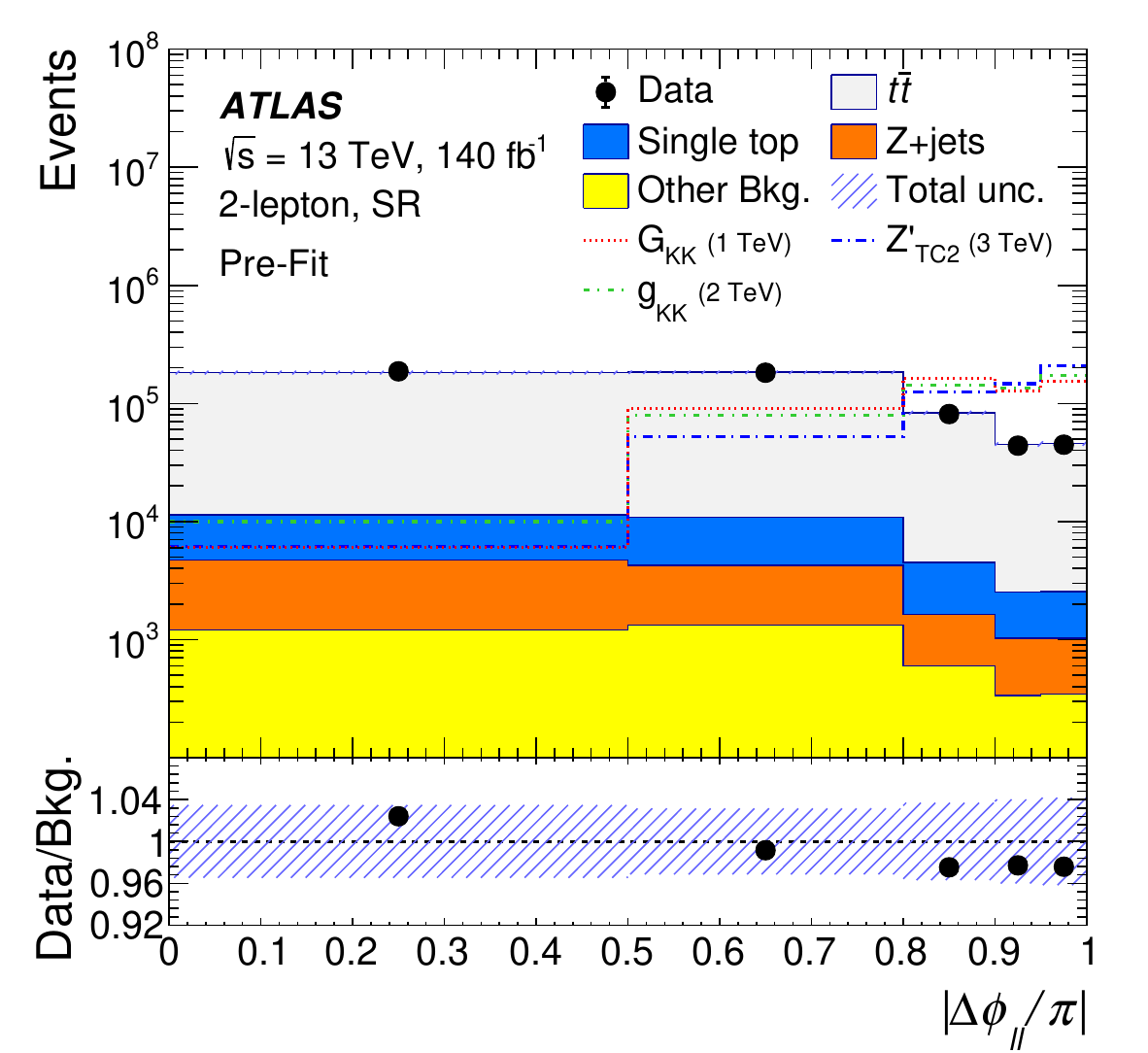}
\label{fig:dPhi:prefit:incl_SR}
}
\subfloat[]{
\includegraphics[width=0.4\textwidth]{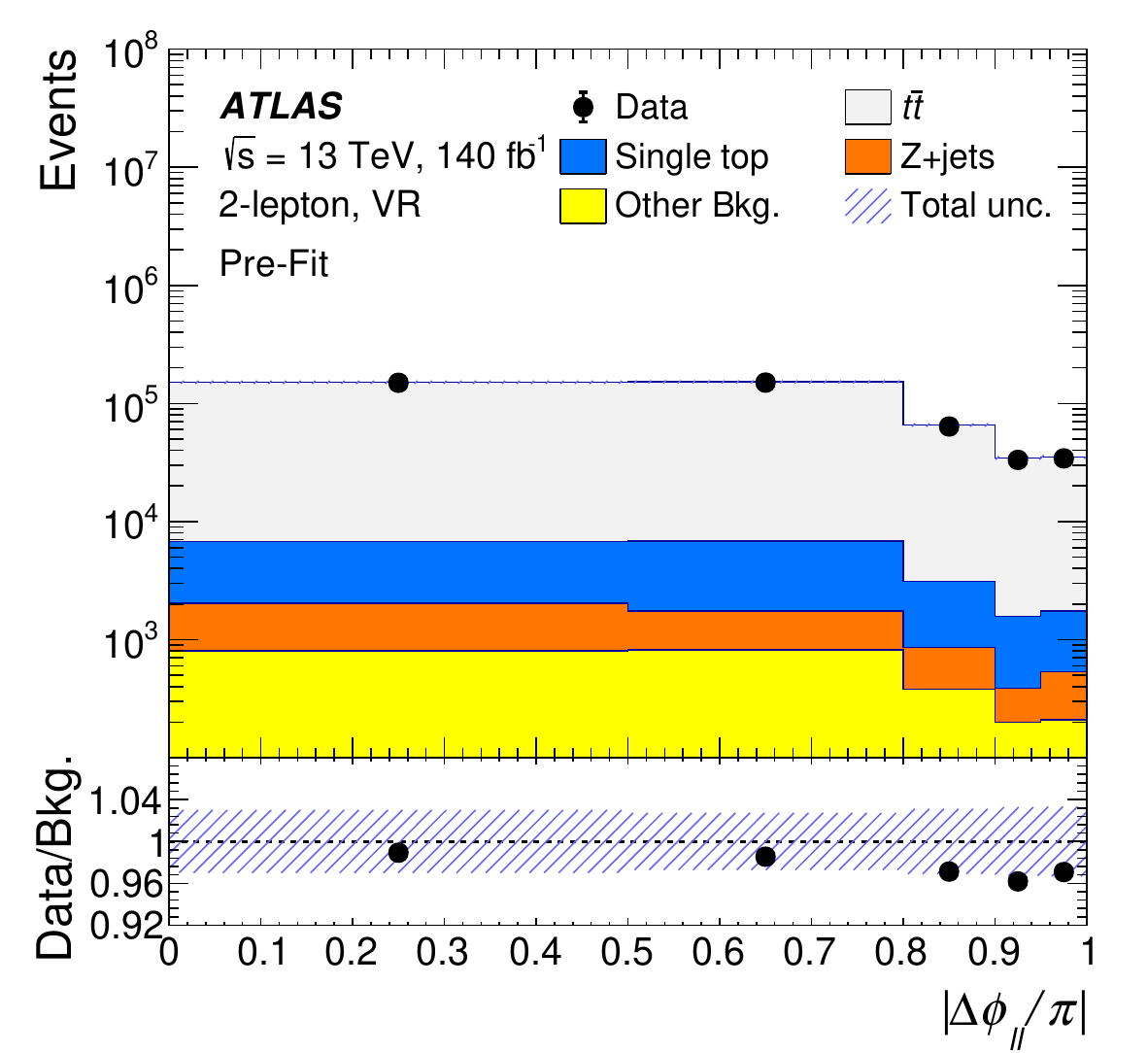}
\label{fig:dPhi:prefit:incl_VR}
} \\
\caption{Distribution of the reconstructed \dPhiPi in \protect\subref{fig:dPhi:prefit:incl_SR} the signal regions and \protect\subref{fig:dPhi:prefit:incl_VR} the validation regions of the 2-lepton channel. Each bin corresponds to one of the signal or validation regions. The distributions and the uncertainty band are obtained before the final fit to the data (pre-fit). ``Other Bkg.'' combines the \ttbarV, \ttbarH, diboson and fakes backgrounds together. In \protect\subref{fig:dPhi:prefit:incl_SR}, a 3~\TeV \zprimeTC signal (blue dashed and dotted line), a 1~\TeV \grav signal (red dotted line) and a 2~\TeV \glu signal (green short-dashed and dotted line) are overlaid for illustration purposes, normalised to the total background. The lower panels show the ratio of the data to the total SM background prediction. The error bands include both the statistical and systematic uncertainties.}
\label{fig:dPhi:prefit:incl}
\end{figure}

In addition to the signal and validation regions defined above, an orthogonal region is used to correct the MC-based modelling of the \Zjets background component. A control region enriched in \Zjets events, \CRZ, is defined by requiring events to satisfy the common selection requirements and additionally contain two same-flavour leptons with opposite charge and with an invariant mass consistent with leptons arising from a \textit{Z} boson decay, \(81<m_{\ell\ell}<101\)~\GeV.


%
\section{Observables}
%

The primary observable used to separate signal from background is the invariant mass of the \ttbar system, \mttbar. Additionally, angular variables sensitive to the \ttbar spin correlations are used in both the 1- and 2-lepton channels to define validation regions for the SM \ttbar background and, in the case of the 2-lepton channel, to increase the sensitivity of the signal regions. The angular variables provide additional discrimination power between the signal process, in which the \ttbar system is produced via the \textit{s}-channel diagram, and the background from SM \ttbar production, which mostly proceeds via the \textit{t}-channel diagram.

\subsection{Observables in the 1-lepton channel}
\label{subsec:reco_1l}
The calculation of \mttbar requires the correct identification and reconstruction of all \ttbar decay products, including the four-momentum of the neutrino from the leptonic decay of one of the two top quarks. In both the resolved and the merged event topologies, the transverse momentum of the neutrino is taken to be the transverse missing momentum in the event, \ptmiss. The four-momentum component longitudinal to the beam axis, \(p_z^\nu\), is calculated from the kinematic constraint that the squared sum of the neutrino and charged lepton four-momenta must yield the squared mass of the \textit{W} boson~\cite{TOPQ-2012-18}. If the resulting equation has exactly one real solution, it is taken as the neutrino longitudinal momentum. If it has no real solutions, the \ptmiss vector is rescaled and rotated in the transverse plane by the minimal amount required to obtain exactly one real solution~\cite{EXOT-2015-04}. This procedure is based on the assumption that the lack of a real solution is caused by a mis-measurement of the transverse missing momentum. If two real solutions are found, in the resolved topology selection, the choice is made with a \(\chi^2\) algorithm, while in the merged topology selection the solution with the smallest absolute value of \(p_z^{\nu}\) is chosen~\cite{EXOT-2015-04}.

\subsubsection{Resolved topology}
\label{subsubsec:reco_1l:resolved}
A \(\chi^2\) minimisation approach is used to select the four jets from the \ttbar decay from all selected \smallR jets and assign them to the leptonically- and hadronically-decaying top quarks, identically to what is done in Ref.~\cite{EXOT-2020-25}. All possible jet permutations are tested, and the one with the lowest \(\chi^2\) is used; if there are two solutions for the neutrino longitudinal momentum, the one that yields the lowest \(\chi^2\) is kept. Any \textit{b}-tagged jet can only be assigned to either of the \textit{b}-quarks produced by the decay of a top quark, reducing the number of permutations to test. The experimental resolution of \mttbar in the resolved signal region is around 12\% for \(\mttFSR = 400\)~\GeV and 10\% for \(\mttFSR = 1000\)~\GeV.

\subsubsection{Merged topology}
For the merged topology, the hadronically-decaying top quark can be straightforwardly identified with the top-tagged \largeR jet that passes the requirements in Section~\ref{sec:objects}. The leptonically-decaying top quark is reconstructed from the selected jet identified with the \textit{b}-jet from the leptonic top-quark decay, the selected lepton, and the reconstructed neutrino (Section~\ref{subsec:selection_1l}). The \mttbar resolution for events in the merged signal region is around 10\% for \(m_{t\bar{t}} > 600\)~\GeV.

In Figure~\ref{fig:reco:1L}, the distributions of the reconstructed \ttbar invariant mass are shown for various representative resonance masses in the three benchmark models described in Section~\ref{sec:models}. At low resonance masses, the \mttbar distributions show a clear peak at the resonance mass. At larger resonance masses, a tail towards masses lower than the generated mass is visible, which arises from off-shell production of the signal and the
growth of low-\textit{x} partons in the proton PDF (where \textit{x} stands for the Bjorken-\textit{x} of the proton)~\cite{TOPQ-2012-18}.
\begin{figure}[ht!]
\centering
\subfloat[]{
\includegraphics[width=0.47\textwidth]{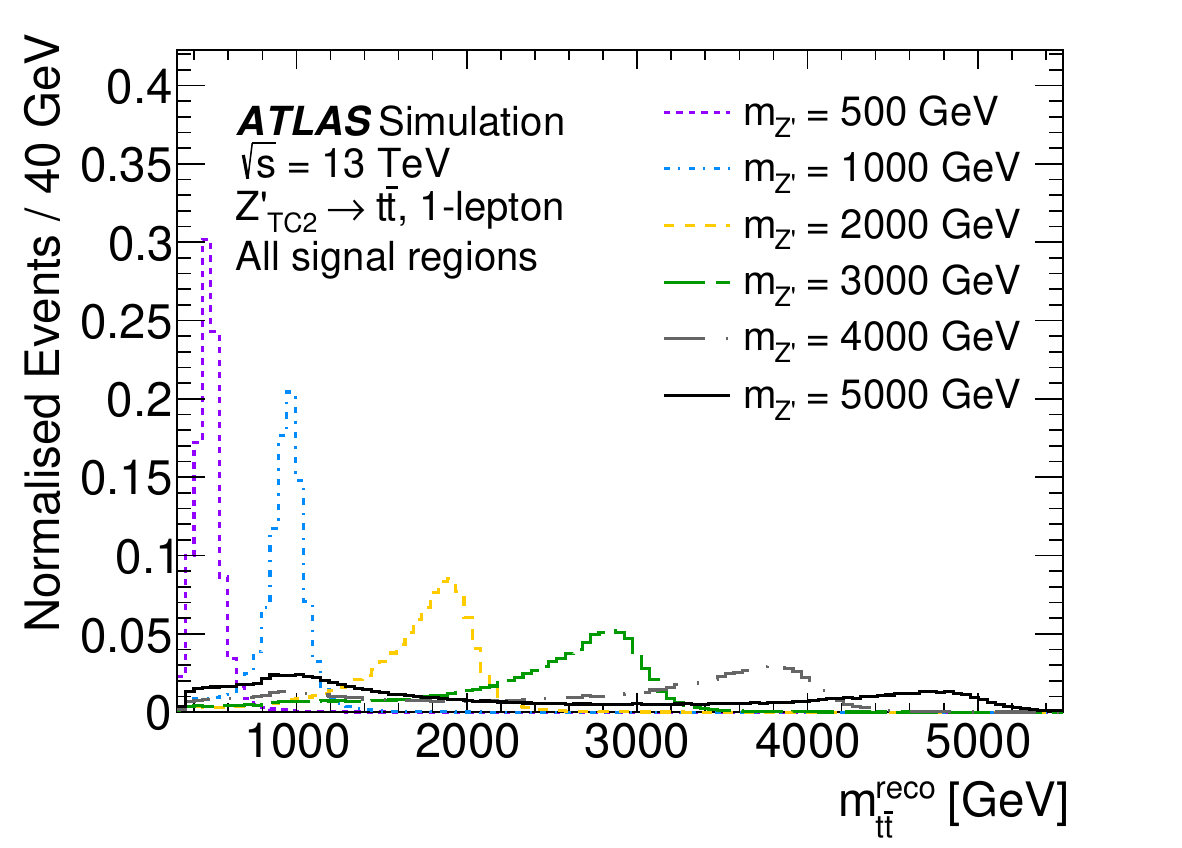}
\label{fig:reco:1L_Zp}
}
\subfloat[]{
\includegraphics[width=0.47\textwidth]{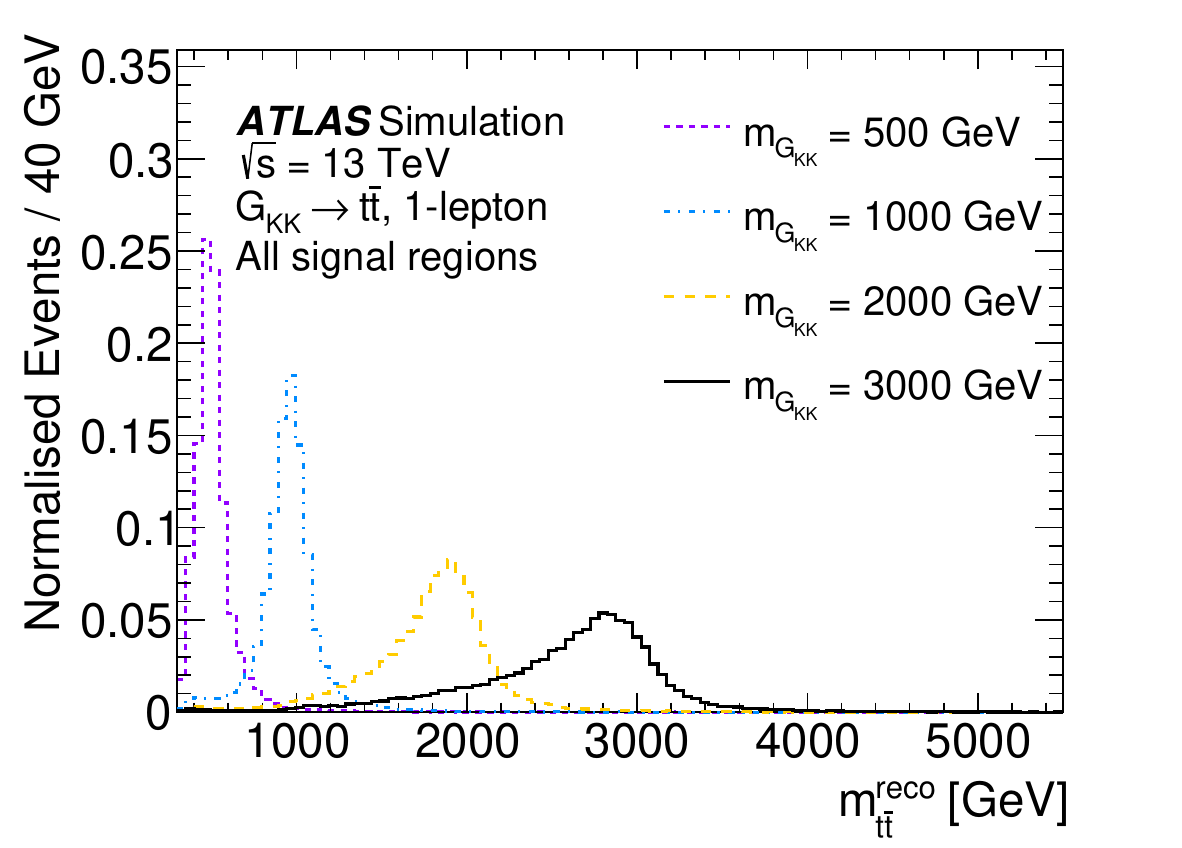}
\label{fig:reco:1L_KKgrav}
} \\
\subfloat[]{
\includegraphics[width=0.47\textwidth]{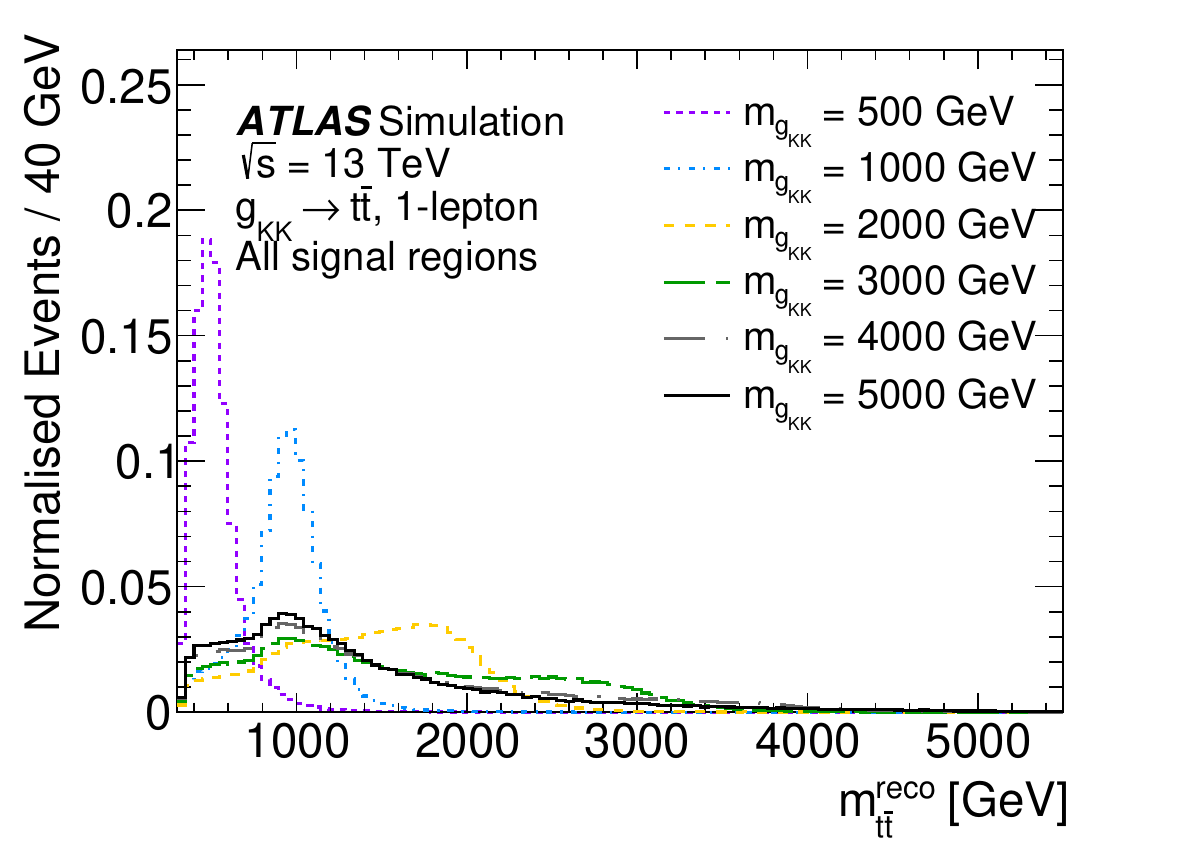}
\label{fig:reco:1L_KKglu}
}
\caption{Signal distributions in \mttbar after the signal region selection of the 1-lepton channel for the \protect\subref{fig:reco:1L_Zp} \zprimeTC, \protect\subref{fig:reco:1L_KKgrav} \grav, and \protect\subref{fig:reco:1L_KKglu} \glu signals for various representative values of the resonance mass. Events from the Resolved \(1b\), Resolved \(2b\), and Merged signal regions are included.}
\label{fig:reco:1L}
\end{figure}

\subsection{Observables in the 2-lepton channel}
The invariant mass of the \ttbar system cannot be unambiguously reconstructed in the 2-lepton channel due to the presence of the two neutrinos from the two leptonically-decaying top quarks. Therefore, instead of the \ttbar invariant mass, the invariant mass of the two selected \textit{b}-jets and the two charged leptons, \mllbb, is used as the discriminating variable.

Figure~\ref{fig:reco:2L} shows the distributions of the reconstructed \mllbb discriminant for the same representative signal masses as shown in the 1-lepton channel. The same trend of low-mass tails for the high-mass signals that were seen in the \mttbar distributions are also visible in the \mllbb distributions.
\begin{figure}[ht!]
\centering
\subfloat[]{
\includegraphics[width=0.47\textwidth]{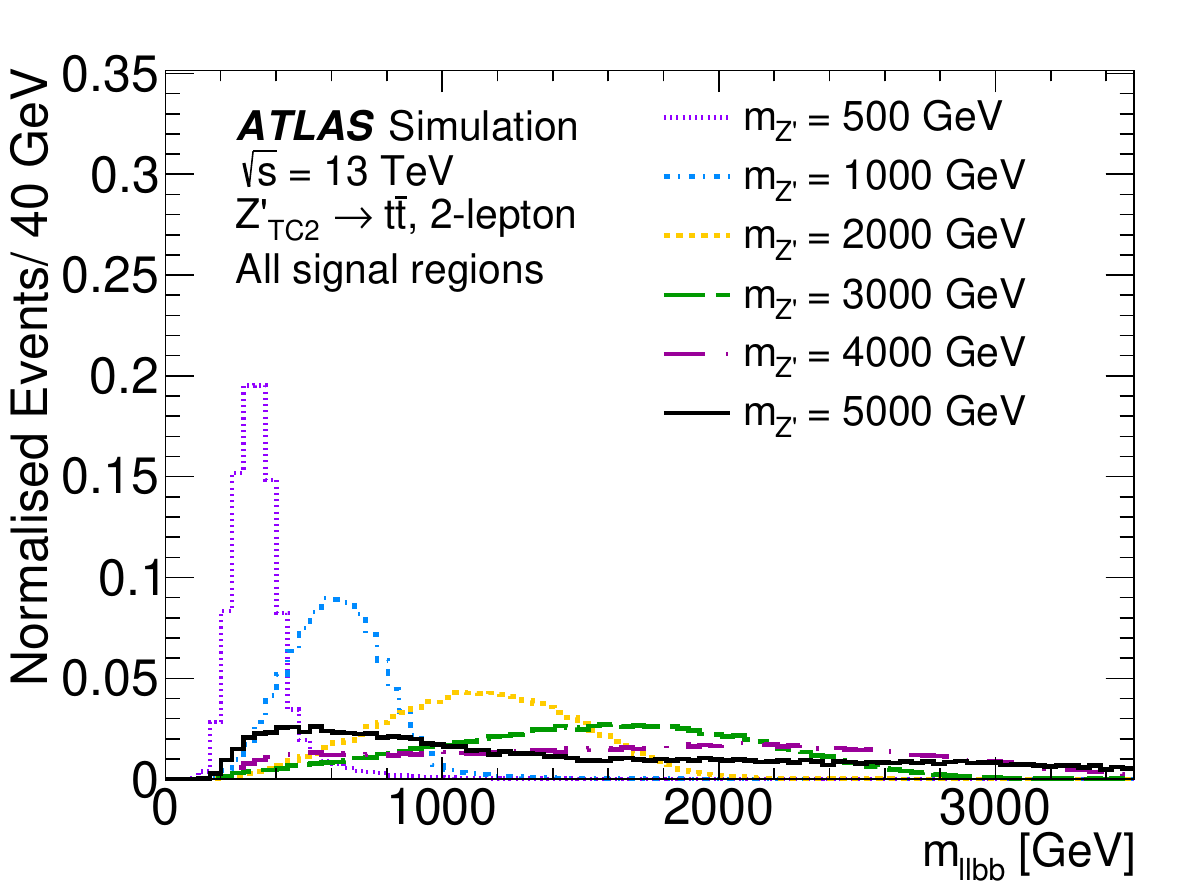}
\label{fig:reco:2L_Zp}
}
\subfloat[]{
\includegraphics[width=0.47\textwidth]{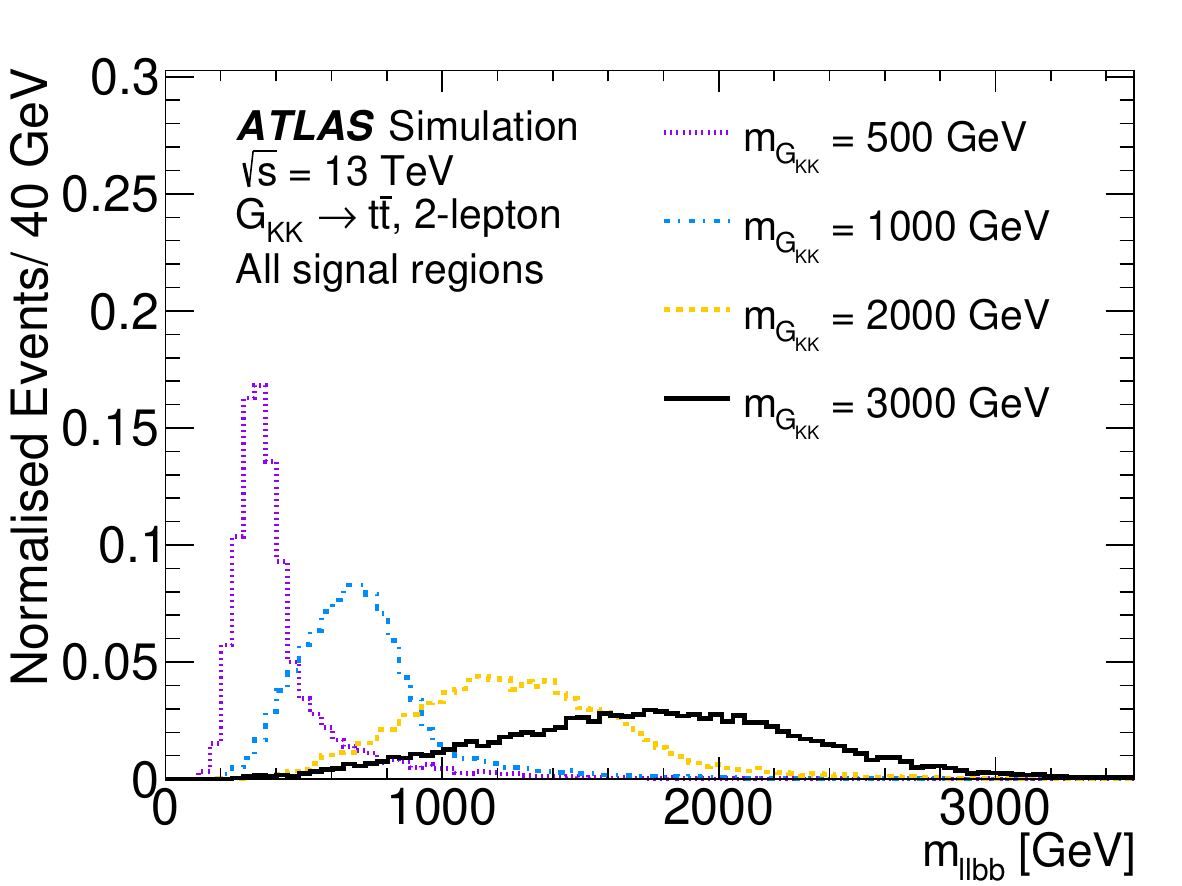}
\label{fig:reco:2L_KKgrav}
} \\
\subfloat[]{
\includegraphics[width=0.47\textwidth]{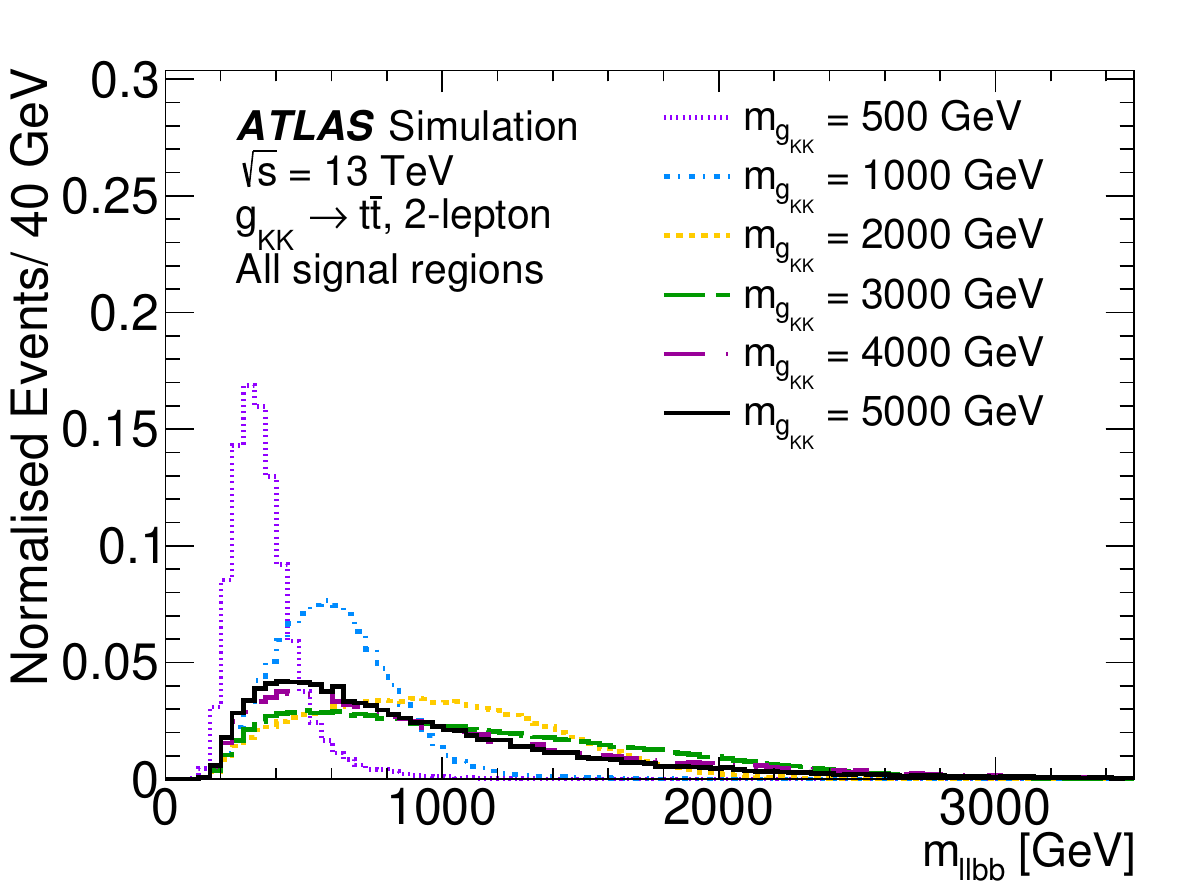}
\label{fig:reco:2L_KKglu}
}
\caption{Signal distributions in \mllbb after the signal region selection of the 2-lepton channel for the \protect\subref{fig:reco:2L_Zp} \zprimeTC, \protect\subref{fig:reco:2L_KKgrav} \grav, and \protect\subref{fig:reco:2L_KKglu} \glu signals for various representative values of the resonance mass. Events from the five signal regions are combined.}
\label{fig:reco:2L}
\end{figure}


%
\section{Data-driven estimation of background contributions}
\label{sec:bkg}
%

\subsection{\Wjets background in the 1-lepton channel}
\label{subsec:wjets_1l}
Scale factors derived from data are applied before the final analysis to correct the normalisation of the \Wjets background in the \SHERPA MC simulation samples, to account for possible mis-modelling of the cross-section times acceptance. The scale factors are determined by comparing the measured \textit{W} boson charge asymmetry in data with that predicted by the simulation. Given that \(pp\) collisions at the LHC result in a charge-asymmetric initial state, the total number of \Wjets events in data in a given signal region, \(N_{W^{+}}+N_{W^{-}}\), can be written as
\begin{equation}
N_{W^{+}}+N_{W^{-}} = \frac{N_{W^{+}}^{\mathrm{MC}}+N_{W^{-}}^{\mathrm{MC}}}{N_{W^{+}}^{\mathrm{MC}}-N_{W^{-}}^{\mathrm{MC}}} (D_{\mathrm{corr+}} - D_{\mathrm{corr-}})\,,
\label{eq:wjets_ca}
\end{equation}
where \(N_{W^{\pm}}^{\mathrm{MC}}\) denotes the number of \Wjets events with a positively- or negatively-charged lepton obtained from the \SHERPA MC simulation. \(D_{\mathrm{corr+(-)}}\) is the number of observed data events with a positively-charged (negatively-charged) lepton. This method relies on the fact that the charge asymmetry obtained in MC simulation is in excellent agreement with the value measured in data~\cite{STDM-2017-13}. Contributions to \(D_{\mathrm{corr+(-)}}\) from charge-asymmetric processes such as single-top, \(WZ\), and \ttbarW production are estimated from MC simulation and are subtracted from the data samples. Contributions from charge-symmetric processes such as \ttbar production and the signal process itself cancel out in the difference on the right-hand side of Eq.~(\ref{eq:wjets_ca}). The final scale factor, \(C_A=1.125\pm0.031\), is then calculated as the ratio of \(N_{W^{+}}+N_{W^{-}}\) evaluated from data to that predicted from the \SHERPA MC simulation. The scale factor is evaluated for the combined \ejets and \mujets events channels in the merged and resolved regions combined; it was verified that the scale factors evaluated from the \ejets and \mujets channels separately are consistent.

The scale factors obtained in the method described above are applied to the single, combined template for the \Wjets background. A 50\% uncertainty is assigned to the normalisation after the \(C_A\) correction to account for additional theory uncertainties related to the \Wjets sample, discussed further in Section~\ref{subsec:unc_mod_nontt}. Furthermore, the fraction of \wHF in the \Wjets nominal template is varied by \(\pm50\%\), with this variation applied as an additional systematic uncertainty to cover any residual mis-modelling arising from the relative flavour contributions~\cite{EXOT-2015-04}.

\subsection{Fakes background in the 1-lepton channel}
\label{subsec:multijet_1l}
The fakes background in the 1-lepton channel consists of events with a non-prompt lepton or a jet mis-reconstructed as a lepton that satisfies the lepton identification and isolation criteria applied in this search (referred to as \textit{tight} lepton requirements in the following). In the \mujets channel, muons arising from semileptonic decays of hadrons constitute the main source of this background. In the \ejets channel, additional fakes arise from events containing jets with a large electromagnetic component, for example from \(\pi^0\rightarrow\gamma\gamma\) decays, or photons mis-identified as isolated electrons. The normalisation, \ttbar invariant mass shape, statistical and systematic uncertainties associated with the fakes background are estimated from data using the \textit{matrix method}. The matrix method used in this search is based on the one used in previous ATLAS \ttbar resonance searches and measurements~\cite{EXOT-2015-04,TOPQ-2012-18} and is the same as that employed in Ref.~\cite{EXOT-2020-25}.

The number of events with leptons satisfying loose lepton identification and isolation criteria~\cite{EGAM-2018-01}, \(N_L\), is defined as
\begin{equation}
N_{\mathrm{L}} = N_{\mathrm{prompt}} + N_{\mathrm{fakes}}\,,
\label{eq:multijet_loose}
\end{equation}
where \(N_{\mathrm{prompt}}\) denotes the number of events with prompt leptons, and \(N_{\mathrm{fakes}}\) denotes the number of events with non-prompt leptons and other sources mis-identified as leptons. The number of events with leptons satisfying the tight lepton identification criteria applied in the signal regions, \(N_T\), can be written as
\begin{equation}
N_{\mathrm{T}} = \epsilon \times N_{\mathrm{prompt}} + f \times N_{\mathrm{fakes}}\,,
\label{eq:multijet_tight}
\end{equation}
where the \textit{real rate}, \(\epsilon\), is an estimator for the probability of a prompt lepton passing the loose lepton identification criteria to also pass the tight critieria, and the \textit{fake rate}, \(f\), is an estimator for the probability that a lepton from other sources passing the loose criteria also passes the tight criteria. The fake rate is estimated from a control region, defined by applying the same selection criteria as the resolved signal region, but with the missing transverse momentum and transverse mass requirements inverted. The veto on events passing the merged topology selection criteria is also not imposed to define this control region, to increase the available data. Contributions from processes leading to prompt leptons are subtracted from the data in this fakes-enriched control region using their predictions from MC simulation. The real rate is determined using SM \ttbar MC samples. Solving Eqs.~(\ref{eq:multijet_loose}) and~(\ref{eq:multijet_tight}) simultaneously for \(N_{\mathrm{prompt}}\) and \(N_{\mathrm{fakes}}\) yields the fakes contribution in the given signal region. The fakes estimate is performed separately for the \ejets and \mujets events.

A smoothing procedure is applied to the fakes background templates in both the resolved and the merged signal regions to mitigate large statistical fluctuations due to the limited data. In the resolved region, a parabolic smoothing algorithm is applied, whereby the input distribution histogram is rebinned to achieve a smooth behaviour. The limited amount of data has a larger effect in the high-\mttbar tails, and the parabolic smoothing used in the resolved regions was observed to be too agressive for the merged topology, removing all entries in the tail of the distribution. Therefore, a custom smoothing procedure, applied separately for the \ejets and \mujets channels, and separately in the signal and validation regions, was designed and implemented for the fakes background in the 1-lepton merged regions. The algorithm uses an analytic function fit to the unbinned templates to smooth the binned template, with care taken to ensure there is continuity in the function when considering the \mttbar-dependent \(\left|\Delta\eta\left(\ell, t^{\mathrm{had}}\right)\right|\) requirement. The systematic uncertainty assigned to this smoothing procedure is described in Section~\ref{subsec:unc_mod_nontt}.

The fakes modelling before applying the smoothing is validated in separate dedicated validation regions, where only one of the \met or \met+\mwt requirements is inverted. These validation regions contain a background composition that is more similar to that in the signal region, while still having an enhanced fakes contribution. Based on the modelling performance observed in these validation regions, a conservative 50\% normalisation uncertainty that is decorrelated across signal regions is assigned to the fakes background, in addition to the uncertainties related to the estimate and smoothing procedures.

\subsection{\Zjets background in the 2-lepton channel}
\label{subsec:zjets_2l}
Data-driven corrections are applied to the \SHERPA MC simulation samples for the \Zjets background in the 2-lepton channel before the final fit to correct the overall normalisation of this background component and the modelling of relevant kinematic variables. The corrections are derived in the \Zjets enriched \CRZ, where disagreement in the modelling of the background prediction obtained with the \SHERPA MC simulated samples is observed compared with data for two kinematic variables in particular: the transverse momentum of the dilepton system, \ptll, and the invariant mass \mllbb. In both cases, the mis-modelling exhibits a linear trend, with good agreement observed for small values of these variables and a linear increase in the difference between data and MC prediction towards higher values. As a result, a reweighting procedure in \mllbb is used to correct the \SHERPA MC samples for the \Zjets background to the data in \CRZ. Each simulated event is assigned a multiplicative correction weight based on its value of \mllbb:
\begin{equation}
w = a \cdot (1 - b \cdot \mllbb)\,.
\label{eq:CRZ_rew}
\end{equation}
The correction factors \(a\) and \(b\) are derived from a maximum-likelihood fit to the \mllbb distributions, taking into account only the statistical uncertainty in the MC predictions. The following values are obtained for the correction factors:
\begin{equation}
a = 1.190 \pm 0.003\,, \quad b = (-1.7 \pm 0.1) \cdot 10^{-4}\;\mathrm{GeV}^{-1}\,. \nonumber
\end{equation}

Good agreement between the data and the SM expectation is observed for all relevant kinematic variables after applying the reweighting correction. To cover the residual non-closure seen in the \Zjets control region after requiring 2 \textit{b}-jets, two uncertainties are introduced. The first is a flat 30\% normalisation uncertainty, and the second is estimated by varying the slope of Eq.~(\ref{eq:CRZ_rew}) by 30\% of its best-fit value and obtaining the reweighted \mllbb distribution in this case. The difference between the nominal and the varied reweighted \mllbb distributions is taken as the systematic uncertainty.


%
\section{Statistical analysis}
\label{sec:stats}
%

The compatibility of the data with the SM prediction is quantified by a binned profile-likelihood fit~\cite{Cowan:2010js,HistFactory} of the expected \mttbar (1-lepton channel) and \mllbb (2-lepton channel) distributions to the observed distributions. The fit is performed simultaneously in the three signal regions of the 1-lepton channel and the five signal regions of the 2-lepton channel. The shape of the binned \mttbar or \mllbb distributions is parameterised in terms of the signal strength, \(\mu\), interpreted as the ratio of the observed and predicted number of signal events. The SM \ttbar normalisation is a free parameter in the fit, while the other simulated background processes are normalised to their theory predictions, with uncertainties as described in Section~\ref{sec:unc}.

Systematic uncertainties, described in Section~\ref{sec:unc}, are taken into account as nuisance parameters (NPs) constrained by Gaussian distributions. The Beeston-Barlow ``lite'' technique~\cite{Barlow:1993dm} is used to incorporate uncertainties due to the limited data available for the MC samples used in the analysis, referred to as \textit{MC statistics}, where one additional NP with a Poissonian constraint is added per bin of the discriminant distribution to the likelihood function. The statistical uncertainty originating from the fakes background estimate is also accounted for in these same bin-by-bin NPs, see Section~\ref{subsec:unc_mod_nontt} below. The combination of the MC statistical uncertainty and the fakes background uncertainty is referred to as \textit{background statistical uncertainties} in the remainder of this paper.


%

%
\section{Systematic uncertainties}
\label{sec:unc}
%

The modelling of signal and background events is affected by various experimental uncertainties related to the reconstruction, identification, and calibration of object and event properties. In addition, uncertainties related to the theory modelling of the simulated backgrounds, most importantly SM \ttbar production, are taken into account. The largest sources of systematic uncertainties in the analysis vary depending on the mass and width of the signal hypothesis being tested, but overall are related to the modelling of the SM \ttbar and the 1-lepton fakes background processes, followed by uncertainties in the jet energy scale and resolution. Table~\ref{tab:syst} in Section~\ref{sec:results} summarises the main sources of uncertainty for three signal hypotheses and their impact on the observed signal strengths.

\subsection{Modelling uncertainties}
\label{subsec:unc_mod}
\subsubsection{Uncertainties in the modelling of the SM \ttbar background}
Modelling uncertainties are considered relating to the background from SM \ttbar production, which is estimated by using MC simulation at NLO+PS precision, and corrected to NNLO-QCD+NLO-EW calculations (see Section~\ref{sec:datamc}). The background prediction for this process is hence subject to uncertainties in the higher-order calculations and uncertainties related to the reweighting procedure, collectively grouped together as \ttbar NNLO uncertainties in Table~\ref{tab:syst}, and uncertainties due to modelling choices made in the NLO+PS simulation. As mentioned earlier, the impact of including a potential quasi-bound state consistent with that observed by CMS was found to be negligible.

The uncertainties in the higher-order calculations are assessed by variations of the renormalisation and factorisation scales, the PDF set, and the EW contribution, which is not present in the NLO+PS prediction. The scale uncertainties are estimated separately for the (anti)top \pT and \mttbar distributions by varying the renormalisation and factorisation scales independently up and down by a factor of two, yielding four independent scale uncertainty components. The uncertainty in the PDF set is taken as the envelope of the intra-PDF uncertainties of the LUXQED PDF set~\cite{Manohar:2016nzj}, which is used to obtain the NNLO-QCD+NLO-EW predictions. The uncertainty in the EW contribution is estimated by comparing the nominal spectra with those obtained with the NNPDF3.0QED PDF set~\cite{Ball:2013,Ball:2014uwa}, which has been found to yield a significantly different prediction for the EW contributions due to a different treatment of the photon PDF compared to the nominal LUXQED PDF set~\cite{Czakon:2017wor}.

Two uncertainties are assigned to the iterative reweighting method. Possible biases due to the order in which the reweighting is applied on the \mttbar and (anti)top-quark \pT distributions is estimated by comparing the nominal reweighted sample with an alternative sample in which the reweighting is applied in a different order. The second uncertainty is derived to account for differences between the definitions of the parton-level top-quark: in the higher-order calculations, it includes at most two real emissions, whereas for the NLO MC generator, any number of real emissions can be included. To account for this, an alternative MC sample is created with the same generators and settings as the nominal NLO MC sample, but without decaying the top quarks and forbidding PS emissions after the first one. The nominal NLO sample is then reweighted to match the parton-level distributions of this modified sample, with the exact same procedure as the reweighting to the higher-order predictions. The difference between this reweighted sample and the un-reweighted sample is taken as the uncertainty in the observables at detector level.

Uncertainties due to the modelling choices made for the NLO+PS MC simulation are estimated by comparing the nominal SM \ttbar prediction to alternative predictions, obtained by either varying parameters within the nominal generators or with alternative MC generators. The alternative MC samples are reweighted to the respective higher-order predictions if the uncertainties they represent do not affect the parton-level variables used for the reweighting, namely \mttbar and the (anti)top-quark \pT. The following uncertainties in this group are:
\begin{itemize}
\item the uncertainty related to the ME-PS matching, estimated by comparing the predictions obtained with the nominal sample with those obtained from an alternative sample generated with the same generator settings as the nominal sample but setting the \(p_{\mathrm{T}}^{\mathrm{hard}}\) parameter\footnote{The \(p_{\mathrm{T}}^{\mathrm{hard}}\) parameter regulates the definition of the vetoed region of the showering to avoid holes or overlaps in the phase space filled by \POWHEG and \PYTHIA.} in \PYTHIA to 1 instead of 0~\cite{ATL-PHYS-PUB-2023-029};
\item the uncertainty related to the choice of the \hdamp parameter, estimated by comparing the predictions of the nominal sample with those obtained with an alternative sample with the \hdamp parameter increased by a factor of 1.5\mtop compared with its nominal value;
\item variations in the ISR ME calculation, which are estimated by varying the factorisation and renormalisation scales independently up and down by a factor of two; and,
\item variations in the FSR, assessed by varying the renormalisation scale for final-state PS emissions up and down by a factor of two.
\end{itemize}
For the cases of the ME-PS matching and the \hdamp uncertainties, the resulting one-sided uncertainties are symmetrised to obtain effective \textit{up} and \textit{down} variations.

The uncertainty associated with variations of the renormalisation scale in the ISR PS is obtained via the corresponding systematic variation in the A14 tune. This variation does not have an explicit higher-order prediction so the nominal reweighting is applied instead, since the uncertainty also does not affect the parton-level variables used for the reweighting.

Finally, several modelling uncertainties are not reweighted at all and compared with the unreweighted nominal sample. These include:
\begin{itemize}
\item the uncertainty related to the choice of the PS and hadronisation model, which is estimated by comparing the nominal predictions with those obtained from a sample generated at NLO in QCD with \POWHEGBOX[v2] with the same PDF set and \hdamp value as in the nominal sample but interfaced with \HERWIG[7.13]~\cite{Bahr:2008pv,Bellm:2015jjp,Bellm:2017jjp}, using the \HERWIG[7.1] default set of tuned parameters~\cite{Bellm:2015jjp,Bellm:2017jjp};
\item the uncertainty related to the \ttbar lineshape, estimated by comparing the predictions obtained with the nominal sample to those obtained with an alternative sample generated with the same setup as the nominal sample but using \MADSPIN to decay the top quarks; and,
\item the uncertainty related to the choice of \mtop, estimated by comparing the nominal sample with \(\mtop=172.5\)~\GeV to two alternative samples with \mtop values of 171.0~\GeV and 174.0~\GeV, respectively, and multiplying the difference by 0.2. This yields an uncertainty of \(\pm0.3\)~\GeV, corresponding ot the current world-average uncertainty in the top-quark mass~\cite{TOPQ-2019-13}.
\end{itemize}
Again, for both the PS uncertainty and the lineshape uncertainty, the resulting one-sided uncertainties are symmetrised to obtain effective \textit{up} and \textit{down} variations.

\subsubsection{Uncertainties in the modelling of the signal}
Theory uncertainties relating to the scale and PDF+\(\alpha_s\) variations are taken into account for the \zprimeTC signal processes. The impact of uncertainties related to the modelling of the signals on the acceptance times efficiency was assessed and found to be negligible compared with the uncertainties from instrumental sources.

\subsubsection{Uncertainties in the modelling of the non-\ttbar\ backgrounds}
\label{subsec:unc_mod_nontt}
The largest background contributions from processes other than SM \ttbar production arise from single-top quark production and, in the 1-lepton channel, from both \Wjets and fakes.

The main uncertainty in the modelling of single-top quark production is determined from a comparison of the diagram removal scheme and the alternative diagram subtraction scheme~\cite{Frixione:2008yi,ATL-PHYS-PUB-2016-020} for the treatment of interference effects and overlap between SM \ttbar and \(Wt\) production. It is estimated by comparing the nominal \(Wt\) samples, generated with the diagram removal scheme, to a set of alternative samples obtained with the diagram subtraction scheme, using the same generator settings as in the nominal single-top samples. As in the case of the SM \ttbar background, an uncertainty related to the choice of the PS and hadronisation model is estimated for the \(Wt\) background component by comparing the nominal prediction to that obtained with the same ME generator settings but interfaced with \HERWIG[7.16], using the \HERWIG[7.1] default set of tuned parameters~\cite{Bellm:2015jjp,Bellm:2017jjp} and the \MMHT[lo] PDF set~\cite{Harland-Lang:2014zoa}. The uncertainty related to the ME-PS matching is estimated by comparing the predictions obtained with the nominal sample to those obtained from an alternative sample generated with the same generator settings as the nominal sample but setting the \(p_{\mathrm{T}}^{\mathrm{hard}}\) parameter in \PYTHIA to 1 instead of 0. Modelling uncertainties related to the choice of the renormalisation and factorisation scales, the levels of ISR and FSR, and the choice of PDF set have also been evaluated but are found to be negligible. The theory uncertainties in the cross-sections for \(Wt\)-, \textit{t}-, and \textit{s}-channel production are applied as pure normalisation uncertainties in the respective components. They are estimated by summing in quadrature the scale and the PDF+\(\alpha_s\) uncertainties in the nominal cross-sections, and amount to \(\pm5.4\%\) for \(Wt\) production~\cite{Kidonakis:2010ux}, \(^{+4.3\%}_{-3.7\%}\) for \textit{t}-channel production~\cite{Kidonakis:2011wy}, and \(^{+4.4\%}_{-4.1\%}\) for \textit{s}-channel production.

For the \Wjets production in the 1-lepton channel, a \(\pm 50\%\) normalisation uncertainty is assigned, and uncertainties related to the choice of scales and PDF set are taken into account as shape-only uncertainties for the MC modelling. To account for any residual mis-modelling of the flavour fractions in the \Wjets background after applying the correction (see Section~\ref{subsec:wjets_1l}), the heavy-flavour fraction of the template is varied up and down by 50\%, consistent with the corrections to the \Wjets template in Ref.~\cite{EXOT-2015-04}. In the 2-lepton channel, no separate uncertainty is assigned to the background from \Wjets production as it is included in the fakes background component, which has a 30\% uncertainty applied to cover the disagreement observed between data and the MC simulation in a fake-enriched control region\footnote{The fake-enriched control region is defined by requiring two signal leptons with the same-sign electric charge and either \(e\mu\) or \(\mu\mu\) events, and that \(\mwt<100\)~\GeV for at least one of the \textit{W} bosons in the event.}.

The systematic uncertainty applied to the \Zjets background in the 2-lepton channel is related to the data-driven reweighting described in Section~\ref{subsec:zjets_2l}. It consists of two components, one affecting the shape and the other affecting the normalisation of this background component. In the 1-lepton channel, where \Zjets production constitutes an even smaller background component, a 30\% normalisation uncertainty is applied, which covers both the \(\pm5\)\% uncertainty in the \Zjets production cross-section and acceptance uncertainties related to possible mis-modellings of higher jet multiplicities.

The uncertainty assigned to the data-driven estimate of the fakes background in the 1-lepton channel consists of three components:
\begin{itemize}
\item a 50\% normalisation uncertainty that is applied separately to the \ejets and \mujets contributions of the fakes estimate;
\item an uncertainty related to the statistical uncertainty in the original Matrix Method estimate, before smoothing; and,
\item an additional systematic uncertainty regulating the shape of the tail of the distribution in the merged regions, obtained from the smoothing procedure.
\end{itemize}
The dominant uncertainty in the fakes estimate is due to the limited data available for the data-driven estimate.

In the case of the remaining small backgrounds, only normalisation uncertainties are applied. For the backgrounds originating from \ttbarZ, \ttbarW, and \ttbarH production, the uncertainties in the respective higher-order cross-sections are taken into account. They amount to \(^{+10\%}_{-12\%}\), \(^{+13\%}_{-12\%}\), and \(^{+6.8\%}_{-9.8\%}\), respectively. A 50\% normalisation uncertainty is applied to the diboson background to take into account any possible mis-modelling in the production of additional jets~\cite{Alwall:2007fs} and heavy-flavour jets~\cite{HIGG-2017-03}.

\subsection{Experimental uncertainties}
The dominant experimental uncertainties affecting the \mttbar and \mllbb spectra are the uncertainties in the jet energy scale (JES) and resolution (JER) of the \smallR jets, for which there are 30 and 13 uncorrelated NPs included in the fit, respectively. Jet mass scale (JMS) uncertainties are also included for \smallR jets but do not have a large impact. The \smallR jet uncertainties are evaluated from a combination of simulations, test-beam data, and in situ measurements~\cite{JETM-2018-05}. Additional contributions from jet flavour composition, punch-through, single-particle response, calorimeter response to different jet flavours and pile-up are also taken into account. Additionally, an uncertainty is assigned to the efficiency of the JVT requirement on jets~\cite{PERF-2014-03}. For \largeR jets, uncertainties in the JES, JER and JMS are derived in similar ways to \smallR jets, using a combination of simulation, test-beam data, and in situ measurements~\cite{JETM-2018-02}. Uncertainties related to jet mass resolution (JMR) are also included for \largeR jets~\cite{JETM-2018-02}.

Further experimental uncertainties in the correction factors for the \textit{b}-tagging efficiencies~\cite{FTAG-2018-01}, as well as the rates of mis-tagging \textit{c}-\cite{FTAG-2020-08} and light-flavour jets~\cite{FTAG-2019-02}, derived from looking at dedicated flavour-enriched samples in data, are applied to the simulated event samples. An additional \textit{b}-tagging efficiency uncertainty is applied for high-momentum jets (\(p_{T}>400\)~\GeV) to account for uncertainties in the modelling of the track reconstruction in high-\pT environments~\cite{ATL-PHYS-PUB-2021-003}. Uncertainties are applied to \largeR jets that satisfy the top-tagging requirements described in Section~\ref{sec:objects}. They are derived from a combination of measurements and account for background efficiency, signal efficiency, high-\pT extrapolation and inefficiency uncertainties. Correlations with the JES and \textit{b}-tagging uncertainties are taken into account~\cite{ATL-PHYS-PUB-2020-017}.

Smaller experimental uncertainties are related to the efficiencies of the lepton identification, isolation, and reconstruction, as well as the lepton energy scale and resolution~\cite{EGAM-2018-01,MUON-2018-03} and to the scale and resolution of the track soft-term in the \met calculation~\cite{JETM-2020-03}. Variations in the reweighting applied to simulated samples to match the expected mean number of interactions observed in each bunch crossing in data are included. They cover the uncertainty in the ratio between the predicted and measured inelastic cross-section. A constant 0.83\% normalisation uncertainty is applied to all signal and background samples, except for fakes and \Wjets that are estimated from data, accounting for the uncertainty in the integrated luminosity.

\subsection{Uncertainty correlation scheme}
\label{sec:uncert_corr}
All experimental uncertainties are treated as fully correlated across samples and the signal regions of the 1- and 2-lepton channels in the final profile-likelihood fit (see Section~\ref{sec:stats}). The modelling uncertainties are treated as uncorrelated between all background samples, but in general correlated across signal regions. However, the uncertainties in the SM \ttbar background related to the modelling of the PS and hadronisation, the PS-ME matching, the lineshape, the FSR and the choice of the \hdamp parameter are treated as uncorrelated between all 3+5 signal regions of the 1- and 2-lepton channels, and separated into their shape and acceptance components. These uncertainties are obtained by comparing the nominal sample to an alternative SM \ttbar sample in each case (\textit{two-point systematic uncertainties}), an approach that typically yields large pre-fit uncertainties that are often highly constrained in a profile-likelihood fit. Keeping these uncertainties uncorrelated between all signal regions is chosen to prevent constraints propagating across regions, thus minimising the overall constraints on these NPs. Additionally, the uncertainties in the SM \ttbar background related to the choice of renormalisation and factorisation scales cannot be assumed to be fully correlated across different kinematic regimes, and are therefore treated as uncorrelated between the resolved and merged signal regions of the 1-lepton channel and between the 1- and 2-lepton channels, although they are treated as correlated across the different \textit{b}-tagging bins of the 1-lepton channel and the different angular bins of the 2-lepton channel.

The uncertainty in the nomalisation factor assigned to the fakes background in the 1-lepton channel is uncorrelated across all three signal regions, and between the \ejets and \mujets channels. The normalisation factor and \(W+\textrm{HF}\) shape factor assigned to the \Wjets background in the 1-lepton channel are also treated as uncorrelated across the three signal regions. All other modelling NPs are treated as correlated across all signal regions.

\FloatBarrier


%

\section{Results}
\label{sec:results}
%

The data are confirmed to be consistent with the background-only hypothesis through a fit of the observed data to the SM prediction with \(\mu=0\) in the eight signal regions, where only the NPs are allowed to vary. The \mttbar distributions in the three signal regions of the 1-lepton channel and the \mllbb distributions in the five signal regions of the 2-lepton channel after the fit to the full 140~\ifb data sample are shown in Figures~\ref{fig:1l_post} and~\ref{fig:2l_post}, respectively. Good compatibility of the observed \mttbar and \mllbb spectra with the post-fit expected spectra is seen in all signal regions, with a goodness-of-fit probability of 95\% obtained from a saturated model test~\cite{Cousins2013GeneralizationOC}.
\begin{figure}[!t]
\centering
\includegraphics[width=0.97\textwidth]{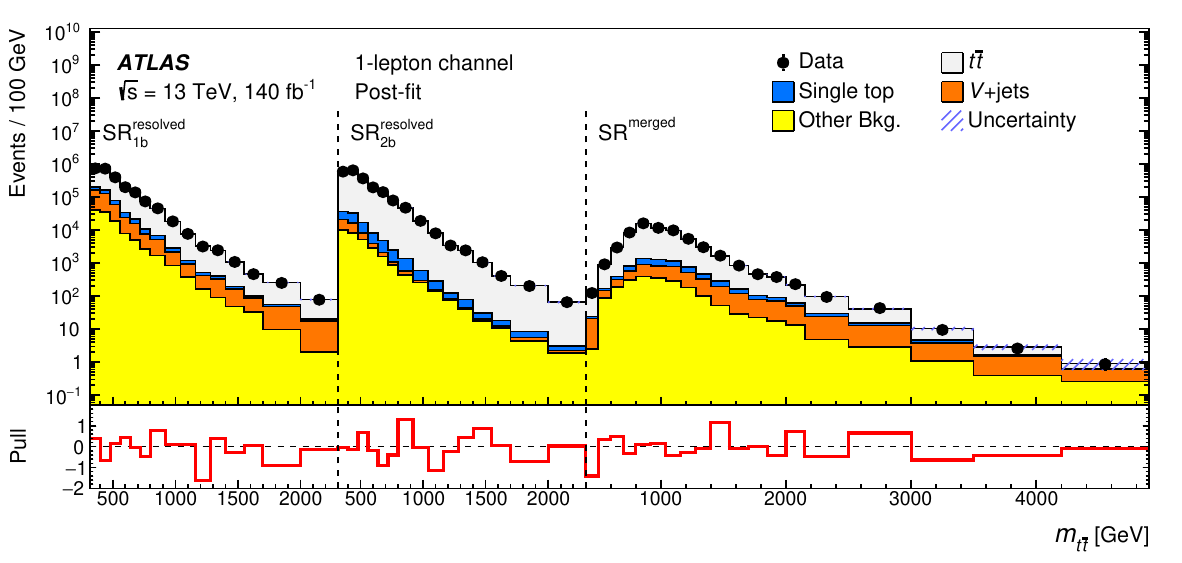}
\caption{Post-fit distributions of the reconstructed \mttbar for the three signal regions of the 1-lepton channel, after performing the profile-likelihood fit under the background-only hypothesis in the 3+5 signal regions of the 1-lepton and 2-lepton channels, respectively. The overflow is added to the final bins. The lower panel shows the pulls from the fit (red line), where the pull in each bin is defined as \((n_{\mathrm{obs}}-n_{\mathrm{pred}})/\sqrt{n_{\mathrm{obs}}}\), and \(n_{\mathrm{obs}}\) and \(n_{\mathrm{pred}}\) refer to the number of observed events in data and the number of predicted events, respectively. ``\Vjets'' refers to the \Zjets and \Wjets backgrounds, whilst ``Other Bkg.'' combines the \ttbarV, \ttbarH, diboson, and fakes backgrounds together.}
\label{fig:1l_post}
\end{figure}
\begin{figure}[!t]
\centering
\includegraphics[width=0.97\textwidth]{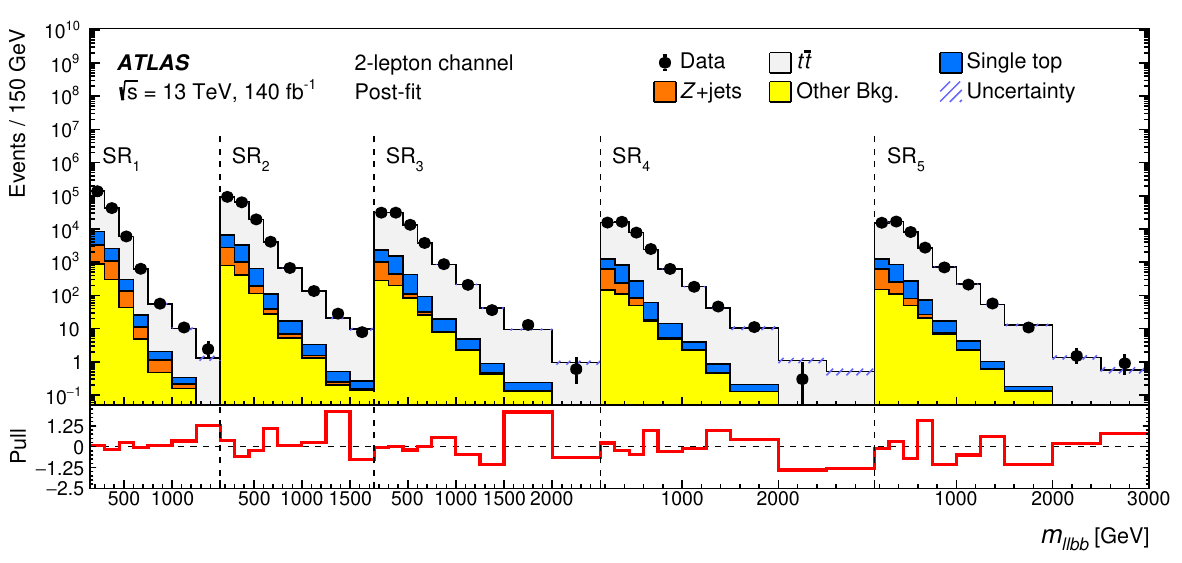}
\caption{Post-fit distributions of the reconstructed \mllbb for the five signal regions of the 2-lepton channel, after performing the profile-likelihood fit under the background-only hypothesis in the 3+5 signal regions of the 1-lepton and 2-lepton channels, respectively. The overflow is added to the final bins. The lower panel shows the pulls from the fit (red line), where the pull in each bin is defined as \((n_{\mathrm{obs}}-n_{\mathrm{pred}})/\sqrt{n_{\mathrm{obs}}}\), and \(n_{\mathrm{obs}}\) and \(n_{\mathrm{pred}}\) refer to the number of observed events in data and the number of predicted events, respectively. ``Other Bkg.'' combines the \ttbarV, \ttbarH, diboson, and fakes backgrounds together.}
\label{fig:2l_post}
\end{figure}

Good modelling of the SM background as a function of \mttbar or \mllbb was validated by applying the post-fit model parameters and constraints from the signal regions in the corresponding validation regions. The results are shown in Figures~\ref{fig:1l_vr} and~\ref{fig:2l_vr} for the 1-lepton and 2-lepton channels, respectively. The correlations of the systematic uncertainties described in Section~\ref{sec:uncert_corr} are maintained, and systematic uncertainties are also correlated between the corresponding signal and validation regions.
\begin{figure}[!t]
\centering
\includegraphics[width=0.97\textwidth]{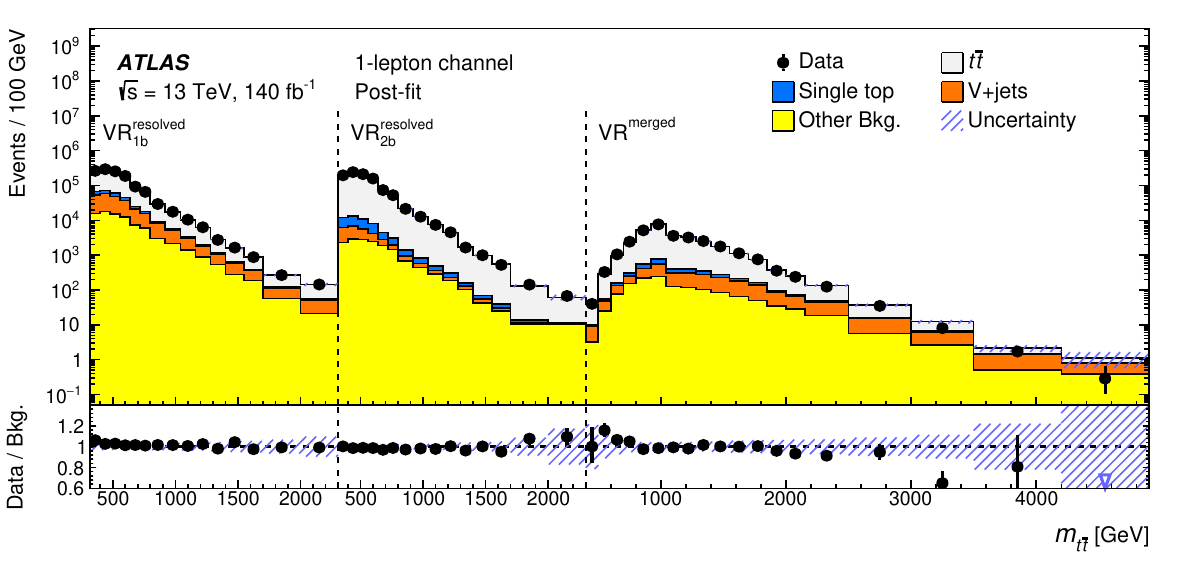}
\caption{Distributions of the reconstructed \mttbar in the three validation regions of the 1-lepton channel, projected from the results of the profile-likelihood fit under the background-only hypothesis in the 3+5 signal regions of the 1-lepton and 2-lepton channels, respectively. The overflow is added to the final bins. In the lower panel, the ratio of the data and the post-fit background in the validation region is shown (data points), as well as the total uncertainty (shaded band). The arrow indicates when the ratio is outside of the range shown. ``\Vjets'' refers to the \Zjets and \Wjets backgrounds, whilst ``Other Bkg.'' combines the \ttbarV, \ttbarH, diboson, and fakes backgrounds together.}
\label{fig:1l_vr}
\end{figure}
\begin{figure}[!t]
\centering
\includegraphics[width=0.97\textwidth]{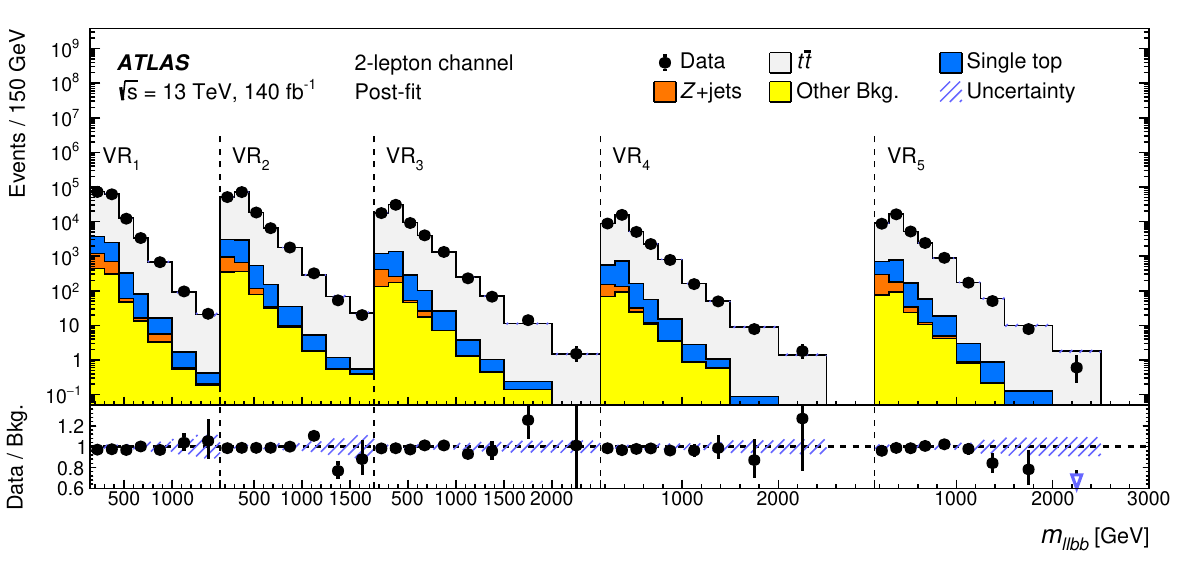}
\caption{Distributions of the reconstructed \mllbb in the five validation regions of the 2-lepton channel, projected from the results of the profile-likelihood fit under the background-only hypothesis in the 3+5 signal regions of the 1-lepton and 2-lepton channels, respectively. The overflow is added to the final bins. In the lower panel, the ratio of the data and the post-fit background in the validation region is shown (data points), as well as the total uncertainty (shaded band). The arrow indicates when the ratio is outside of the range shown. ``Other Bkg.'' combines the \ttbarV, \ttbarH, diboson and fakes backgrounds together.}
\label{fig:2l_vr}
\end{figure}

\subsection{Uncertainty impact}
The relative importance of the different categories of systematic uncertainties is quantified via their post-fit impact on the observed signal strength for three representative signal hypotheses. The results are shown for one narrow, low-mass \zprimeTC signal and one wide, low-mass \glu signal, both of which are fully contained in the resolved signal regions, and one narrow, high-mass \zprimeTC signal, which is contained mostly in the merged signal region. The \grav signals show the same trends as the \zprimeTC, due to the similar widths and similar reconstructed \mttbar shapes; therefore, the discussion below on the \zprimeTC uncertainties also applies to the \grav signals.

The impact of each NP on the signal strength is evaluated from the covariance matrix of the fit, and the impact of a group of NPs is obtained by summing in quadrature the impacts of all NPs in the respective category. Similarly, the total systematic uncertainty is obtained by summing in quadrature the impacts of all relevant NP groups. The statistical uncertainty is calculated by subtracting in quadrature the total systematic uncertainty from the total uncertainty in the signal strength. Table~\ref{tab:syst} summarises the impacts, where the fractional contribution to the total uncertainty in the observed signal strength of each category is given. The observed signal strength and the corresponding absolute uncertainty are shown in the top row.

Of the systematic uncertainties, the dominant impact on the post-fit result depends on the signal hypothesis being tested; however, the \ttbar background modelling uncertainties usually have relatively large impacts for all mass points. The largest experimental systematic uncertainties are those related to the jet energy scale and jet energy resolution of the \smallR jets for the low-mass signals and to top-tagging for the high-mass signals, followed by flavour-tagging related systematic uncertainties. Besides the systematic uncertainties, the NPs related to the statistical uncertainty from the fakes background, included in ``Modelling: statistical uncertainty'' in Table~\ref{tab:syst}, are also important, and for some mass points, these are among the parameters with the highest impact. For the highest mass regions, this is directly due to limited data. On the other hand, for the low-mass, narrow signals, the NPs related to the statistical uncertainty in the fakes are highly-correlated with the parameter of interest. This is because these NPs are completely uncorrelated across bins, and the width of the narrow signals is similar to the bin-width used. The total uncertainty in this analysis is still predominantly statistically dominated for the narrow resonances; however, for the wide resonances, the systematic uncertainties are comparable or even larger than the statistical uncertainties.
\begin{table}[t]
\begin{center}
\caption{Post-fit fractional contributions of different uncertainty categories to the total uncertainty in the observed signal strength, \(\mu\), as determined in the combined fit to the 1- and 2-lepton channels for three representative signal hypotheses: a 500~\GeV \zprimeTC signal, a 3~\TeV \zprimeTC signal, and a 500~\GeV \glu signal. The observed value of \(\mu\) and the corresponding absolute uncertainty are also shown. The details of how the impact of each NP on the signal strength is evaluated, as well as how the total systematic and statistical uncertainties are calculated, are given in the text. ``Modelling: statistical uncertainty'' refers to the contribution from the statistical uncertainties of the MC simulated samples and the data-driven background estimate for the 1-lepton fakes background. ``Modelling: signal'' refers to the theory uncertainties on the \zprimeTC signal; these uncertainties are not considered for the \grav and \glu signals. The individual components do not necessarily add up in quadrature to the total due to rounding.}
\label{tab:syst}
\begin{tabular}{l|SSS}
\toprule
& \textbf{\(m_{\zprimeTC}=500\)~\GeV} & \textbf{\(m_{\zprimeTC}=3\)~\TeV} & \textbf{\(m_{\glu}=1\)~\TeV} \\
& \multicolumn{1}{c}{\(\mu=0.019\pm0.027\)} & \multicolumn{1}{c}{\(\mu=0.06\pm0.31\)} & \multicolumn{1}{c}{\(\mu=0.002\pm0.018\)} \\
\midrule
Uncertainty component   & \multicolumn{3}{c}{Fractional contribution [\%]} \\
\midrule
\textbf{Experimental}                           & \bfseries 27 & \bfseries 21 & \bfseries 37 \\
\SmallR jets                                    & 25 & 12 & 31 \\
\LargeR jets 	                                & 1.6 & 3.8 & 5.8 \\
Flavour tagging	                                & 10 & 6.2 & 8.9 \\
Top-tagging	                                    & 2.8 & 15 & 13 \\
Leptons		                                    & 4.1 & 4.8 & 11 \\
Other               	                        & 3.9 & 2.6 & 6.9 \\
\midrule
\textbf{Modelling: SM \ttbar}                   & \bfseries 26 & \bfseries 48 & \bfseries 46 \\
\ttbar NNLO	                                    & 11 & 7.9 & 13 \\
\ttbar lineshape                                & 14 & 19 & 15 \\
\ttbar ME-PS ($p_{\mathrm{T}}^{\mathrm{hard}}$) & 11 & 6.8 & 17 \\
\ttbar ME-PS (\hdamp)	                        & 9.5 & 10 & 14 \\
\ttbar ISR\& FSR                                & 9.7 & 13 & 29 \\
\ttbar PS	                                    & 8.2 & 40 & 20 \\
\midrule
\textbf{Modelling: signal}                       & \bfseries 1.0 & \bfseries 0.78 & \textbf{--}  \\
\midrule
\textbf{Modelling: other}                       & \bfseries 21  & \bfseries 20 & \bfseries 31 \\
Fakes	                                        & 17 & 5.1 & 10 \\
\Wjets                                      	& 9.0 & 18 & 6.8 \\
\Zjets                                      	& 1.9 & 0.31 & 3.1 \\
Other                                           & 9.2 & 3.0 & 28 \\
\midrule
\textbf{Modelling: statistical uncertainty}     & \bfseries 48 & \bfseries 49 & \bfseries 33 \\
\bottomrule
\textbf{Total systematic uncertainty}           & \bfseries 65 & \bfseries 74 & \bfseries 74 \\
\textbf{Total statistical uncertainty}          & \bfseries 76 & \bfseries 67 & \bfseries 67 \\
\bottomrule
\end{tabular}
\end{center}
\end{table}

\subsection{Constraints on selected benchmark models}
No significant excesses are observed across the entire \mttbar or \mllbb spectra; therefore, upper limits at the 95\% confidence level (CL) are derived on the production cross-section times branching ratio to \ttbar for the resonances predicted in the benchmark scenarios described in Section~\ref{sec:models}. The upper limits are calculated using the \CLs frequentist formalism~\cite{Read:2002hq} and the approximate limits for intermediate masses are interpolated between the tested mass hypotheses. For the \grav signal, the detector resolution is finer than the grid spacing, so the interpolation is only for display purposes.

The observed and expected exclusion limits are shown in Figure~\ref{fig:observed_limits} for the \zprimeTC, \glu and \grav signal hypotheses. The observed limits are in good agreement with the expected limits. Upper limits on the cross-section are extracted to be 3.6~pb, 1.8~pb, and 6.4~pb for a 500~\GeV \zprimeTC, \grav, and \glu signal, respectively. The results in this paper are the best limits on the cross-section times branching fraction to \ttbar for the \zprimeTC signal for \(\mttbar<1.5\)~\TeV, and the best across the full \mttbar range considered for the \grav and \glu signals to date. \zprimeTC signals with a width of 3\% (1.2\%) and \glu signals are excluded up to masses of 4.2~\TeV (3.65~\TeV) and 4.1~\TeV, respectively, while the \grav is excluded up to 1.3~\TeV. The \glu mass exclusions set the strongest limits on this benchmark model to date. The \zprimeTC mass exclusions significantly increase the previously excluded ATLAS limits in the 1-lepton channel on 36~\ifb of data; however, the ATLAS search in fully hadronic final states still provide the most stringent mass limits on these models~\cite{EXOT-2018-48}. The mass limit on the \grav signal is the most stringent limit to date in the \ttbar decay channel, although complementary searches in diboson final states do provide stronger mass exclusions.

\begin{figure}[!ht]
\centering
\subfloat[]{
\includegraphics[width=0.45\textwidth]{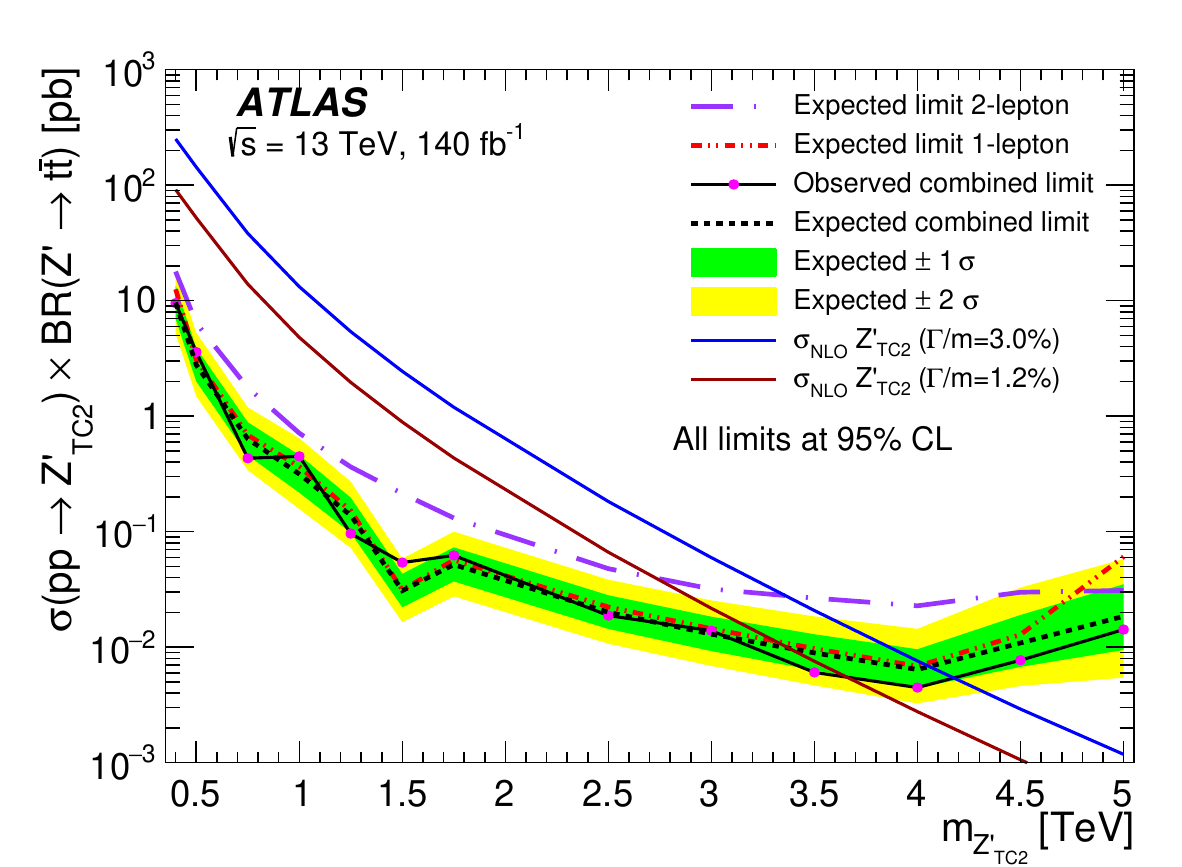}
\label{fig:observed_ZPrime_Limit}
}
\subfloat[]{
\includegraphics[width=0.45\textwidth]{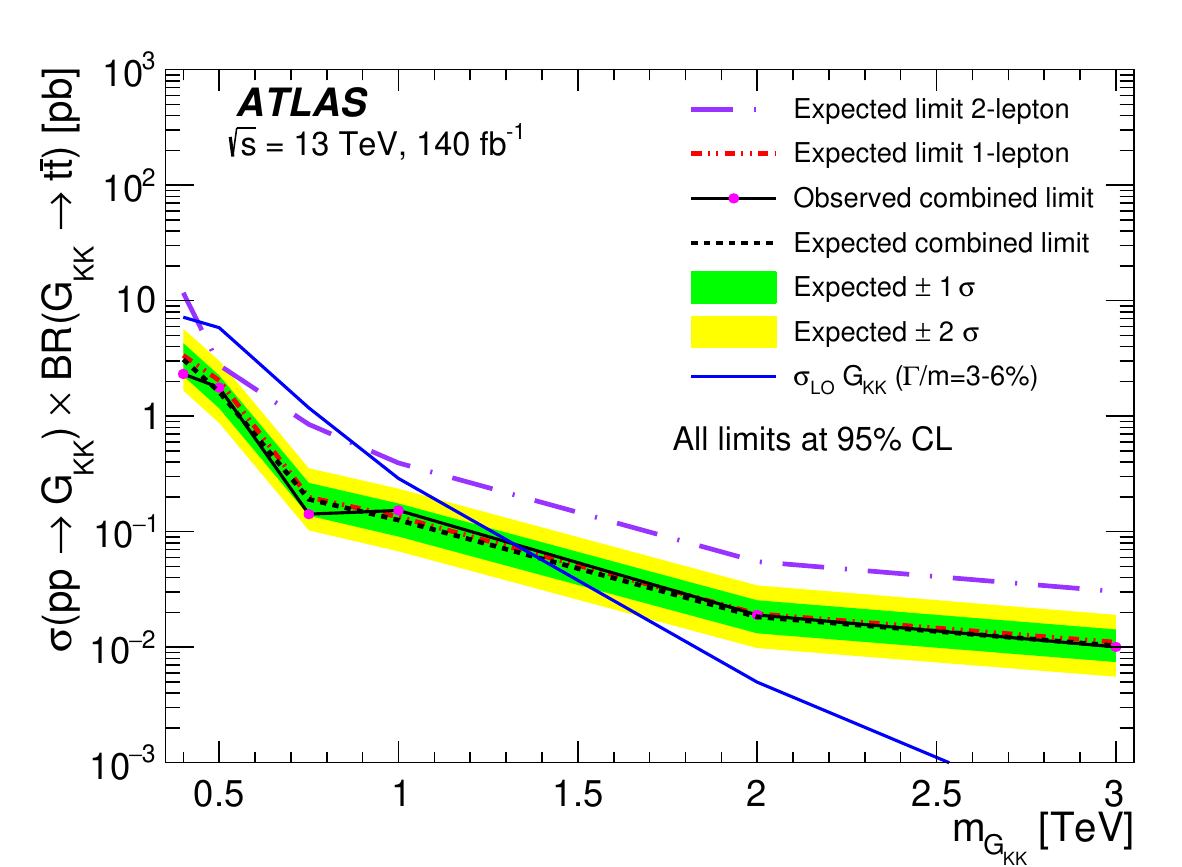}
\label{fig:observed_KKGrav_Limit}
} \\
\subfloat[]{
\includegraphics[width=0.45\textwidth]{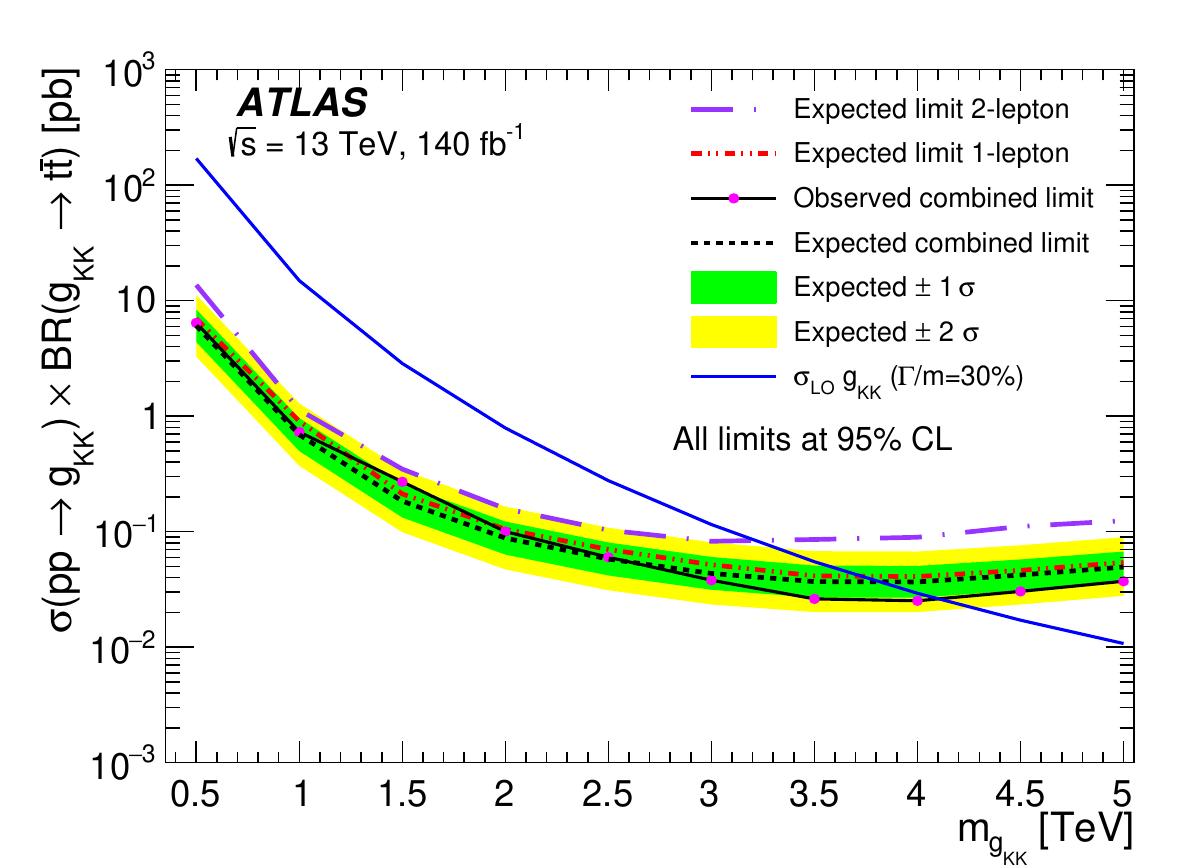}
\label{fig:observed_KKg_Limit}
}
\caption{Observed (pink data points) and expected (black dashed line) upper limits on the signal cross-section times branching ratio to \ttbar for the \protect\subref{fig:observed_ZPrime_Limit} \zprimeTC, \protect\subref{fig:observed_KKGrav_Limit} \grav, and \protect\subref{fig:observed_KKg_Limit} \glu signals at 95\% CL. Approximate limits for intermediate masses are interpolated between the tested mass hypotheses. For the \grav signal in \protect\subref{fig:observed_KKGrav_Limit}, the detector resolution is finer than the grid spacing, so the interpolation is only for display purposes. The theory predictions for the production cross-section times branching ratio at the corresponding masses are also shown in \protect\subref{fig:observed_ZPrime_Limit} for a \zprimeTC with a width of 3\% (blue line) and a \zprimeTC with a width of 1.2\% (dark red line), in \protect\subref{fig:observed_KKGrav_Limit} a \grav with a width ranging from 3\%--6\%, and in \protect\subref{fig:observed_KKg_Limit} a \glu with a width of 30\%. Also shown is the relative contribution of the 1-lepton (red dashed and dotted line) and 2-lepton (purple long-dashed and dotted line) channels to the expected combination limits.}
\label{fig:observed_limits}
\end{figure}


%
%
\FloatBarrier

\section{Conclusion}
%
%

The search presented in this paper targets heavy resonances decaying into top--antitop-quark pairs using final states with exactly one or two leptons. The analysis is performed using 140~\ifb of proton--proton collision data recorded by the ATLAS detector at the LHC at a centre-of-mass energy of \sqrtS. The search covers resonance masses in the range of 0.4~\TeV to 5.0~\TeV and considers both narrow and broad resonances, relative to the detector resolution.

No significant deviation from the Standard Model background expectation is observed across the probed invariant mass spectrum. As a result, upper limits are set on the production cross-section times branching fraction for three benchmark scenarios. For \(\mttbar<1.5\)~\TeV, the search sets the most stringent constraints to date on narrow \ttbar resonances, excluding, for instance, values of the production cross-section times branching fraction greater than 3.6~pb for a \zprimeTC with a mass of 500~\GeV.

The \zprimeTC signal with a relative width of 3\% (1.2\%) is excluded for masses below 4.2~\TeV (3.65~\TeV). These limits are slightly weaker than those obtained in the ATLAS analysis targeting fully-hadronic final states, which excluded masses up to 4.7~\TeV using the same data, yet the present results offer complementary sensitivity in alternative \ttbar decay channels.

Furthermore, the analysis yields the strongest exclusions to date for the \glu signal with a relative width of 30\%, excluding those with a mass below 4.1~\TeV. The constraints on \grav signals are the strongest to date in the studied decay channels, ruling out masses below 1.3~\TeV and significantly extending the reach of previously published results in the \ttbar decay channel.

Overall, the results reported in this study enhance existing exclusion limits across a broad range of models predicting heavy resonances decaying into top-quark pairs.


%
\section*{Acknowledgements}

%
%

%
%

We thank CERN for the very successful operation of the LHC and its injectors, as well as the support staff at
CERN and at our institutions worldwide without whom ATLAS could not be operated efficiently.

The crucial computing support from all WLCG partners is acknowledged gratefully, in particular from CERN, the ATLAS Tier-1 facilities at TRIUMF/SFU (Canada), NDGF (Denmark, Norway, Sweden), CC-IN2P3 (France), KIT/GridKA (Germany), INFN-CNAF (Italy), NL-T1 (Netherlands), PIC (Spain), RAL (UK) and BNL (USA), the Tier-2 facilities worldwide and large non-WLCG resource providers. Major contributors of computing resources are listed in Ref.~\cite{ATL-SOFT-PUB-2026-001}.

We gratefully acknowledge the support of ANPCyT, Argentina; YerPhI, Armenia; ARC, Australia; BMWFW and FWF, Austria; ANAS, Azerbaijan; CNPq and FAPESP, Brazil; NSERC, NRC and CFI, Canada; CERN; ANID, Chile; CAS, MOST and NSFC, China; Minciencias, Colombia; MEYS CR, Czech Republic; DNRF and DNSRC, Denmark; IN2P3-CNRS and CEA-DRF/IRFU, France; SRNSFG, Georgia; BMFTR, HGF and MPG, Germany; GSRI, Greece; RGC and Hong Kong SAR, China; ICHEP and Academy of Sciences and Humanities, Israel; INFN, Italy; MEXT and JSPS, Japan; CNRST, Morocco; NWO, Netherlands; RCN, Norway; MNiSW, Poland; FCT, Portugal; MNE/IFA, Romania; MSTDI, Serbia; MSSR, Slovakia; ARIS and MVZI, Slovenia; DSI/NRF, South Africa; MICIU/AEI, Spain; SRC and Wallenberg Foundation, Sweden; SERI, SNSF and Cantons of Bern and Geneva, Switzerland; NSTC, Taipei; TENMAK, T\"urkiye; STFC/UKRI, United Kingdom; DOE and NSF, United States of America.

Individual groups and members have received support from BCKDF, CANARIE, CRC and DRAC, Canada; CERN-CZ, FORTE and PRIMUS, Czech Republic; COST, ERC, ERDF, Horizon 2020 and Marie Sk{\l}odowska-Curie Actions, European Union; Investissements d'Avenir Labex, Investissements d'Avenir Idex and ANR, France; DFG and AvH Foundation, Germany; Herakleitos, Thales and Aristeia programmes co-financed by EU-ESF and the Greek NSRF, Greece; BSF-NSF and MINERVA, Israel; NCN and NAWA, Poland; La Caixa Banking Foundation, CERCA and AGAUR programs from Generalitat de Catalunya and PROMETEO and GenT Programmes Generalitat Valenciana, Spain; G\"{o}ran Gustafssons Stiftelse, Sweden; The Royal Society and Leverhulme Trust, United Kingdom; Eric and Wendy Schmidt Fund for Strategic Innovation, United States of America.

In addition, individual members wish to acknowledge support from Chile: Agencia Nacional de Investigaci\'on y Desarrollo (FONDECYT 1230812, FONDECYT 1230987, FONDECYT 1240721, FONDECYT 1240864, FONDECYT 3240661); China: Chinese Ministry of Science and Technology (MOST-2023YFA1605700), National Natural Science Foundation of China (NSFC-12175119); Czech Republic: Czech Science Foundation (GACR - 24-11373S), Ministry of Education Youth and Sports (ERC-CZ-LL2327, FORTE CZ.02.01.01/00/22\_008/0004632), PRIMUS Research Programme (PRIMUS/21/SCI/017); EU: H2020 European Research Council (ERC - 101002463); European Union: European Research Council (BARD No. 101116429, ERC - 948254, ERC 101089007), European Regional Development Fund (HE COFUND GA No.101081355, ERDF), European Union, Future Artificial Intelligence Research (FAIR-NextGenerationEU PE00000013), Marie Sklodowska-Curie Actions (ERC Horizon-2020, GA 956086, GAP-101168829); France: Agence Nationale de la Recherche (ANR-21-CE31-0013, ANR-21-CE31-0022, ANR-22-EDIR-0002, ANR-24-CE31-0504-01); Germany: Deutsche Forschungsgemeinschaft (DFG - 469666862, DFG - CR 312/5-2); China: Research Grants Council (GRF); Italy: Ministero dell'Università e della Ricerca (NextGenEU 153D23001490006 M4C2.1.1, NextGenEU I53D23000820006 M4C2.1.1, NextGenEU I53D23001490006 M4C2.1.1, SOE2024\_0000023); Japan: Japan Society for the Promotion of Science (JSPS KAKENHI  JP25H0063, JSPS KAKENHI JP22H01227, JSPS KAKENHI JP22H04944, JSPS KAKENHI JP22KK0227, JSPS KAKENHI JP24K23939, JSPS KAKENHI JP24KK0251, JSPS KAKENHI JP25H00650, JSPS KAKENHI JP25H01291, JSPS KAKENHI JP25K01011, JSPS KAKENHI JP25K01023); Norway: Research Council of Norway (RCN-314472); Poland: Ministry of Science and Higher Education (IDUB AGH, POB8, D4 no 9722), Polish National Science Centre (NCN OPUS 2023/51/B/ST2/02507, NCN OPUS nr 2022/47/B/ST2/03059, UMO-2019/34/E/ST2/00393, UMO-2021/42/E/ST2/00350, UMO-2022/47/O/ST2/00148, UMO-2023/49/B/ST2/04085, UMO-2023/51/B/ST2/00920, UMO-2024/53/N/ST2/00869); Portugal: Foundation for Science and Technology (FCT); Spain: Agència de Gestió d'Ajuts Universitaris i de Recerca. (AGAUR - 2023 BP 00141), Generalitat Valenciana (ASFAE/2022/008), Ministry of Science and Innovation (RYC2019-028510-I, RYC2020-030254-I, RYC2021-031273-I, RYC2022-038164-I), Ministerio de Ciencia, Innovación y Universidades/Agencia Estatal de Investigaci\'on (EU NextGenerationEU (PRTR-C17.I1), PID2022-142604OB-C22); Sweden: Carl Trygger Foundation (Carl Trygger Foundation CTS 22:2312), Swedish Research Council (Swedish Research Council 2023-04654, VR 2021-03651, VR 2022-03845, VR 2022-04683, VR 2023-03403, VR 2024-05451), Knut and Alice Wallenberg Foundation (KAW 2018.0458, KAW 2022.0358, KAW 2023.0366); Switzerland: Swiss National Science Foundation (SNSF - PCEFP2\_194658); United Kingdom: The Binks Trust, Royal Society (NIF-R1-231091); United States of America: U.S. Department of Energy (ECA DE-AC02-76SF00515), Neubauer Family Foundation.

%
%


%

%
%
%
%
%

%
%
%
\printbibliography
\clearpage
 
\begin{flushleft}
\hypersetup{urlcolor=black}
{\Large The ATLAS Collaboration}

\bigskip

\AtlasOrcid[0000-0002-6665-4934]{G.~Aad}$^\textrm{\scriptsize 103}$,
\AtlasOrcid[0000-0001-7616-1554]{E.~Aakvaag}$^\textrm{\scriptsize 17}$,
\AtlasOrcid[0000-0002-5888-2734]{B.~Abbott}$^\textrm{\scriptsize 121}$,
\AtlasOrcid[0000-0002-0287-5869]{S.~Abdelhameed}$^\textrm{\scriptsize 84b}$,
\AtlasOrcid[0000-0002-1002-1652]{K.~Abeling}$^\textrm{\scriptsize 55}$,
\AtlasOrcid[0000-0001-5763-2760]{N.J.~Abicht}$^\textrm{\scriptsize 49}$,
\AtlasOrcid[0000-0002-8496-9294]{S.H.~Abidi}$^\textrm{\scriptsize 30}$,
\AtlasOrcid[0009-0003-6578-220X]{M.~Aboelela}$^\textrm{\scriptsize 45}$,
\AtlasOrcid[0000-0002-9987-2292]{A.~Aboulhorma}$^\textrm{\scriptsize 36e}$,
\AtlasOrcid[0000-0001-5329-6640]{H.~Abramowicz}$^\textrm{\scriptsize 155}$
% ...
\bigskip
\\

% ...
$^{17}$Department for Physics and Technology, University of Bergen, Bergen; Norway.\\
% ...
$^{30}$Physics Department, Brookhaven National Laboratory, Upton NY; United States of America.\\
% ...
$^{36}$$^{(a)}$Facult\'e des Sciences Ain Chock, Universit\'e Hassan II de Casablanca;$^{(b)}$Facult\'{e} des Sciences, Universit\'{e} Ibn-Tofail, K\'{e}nitra;$^{(c)}$Facult\'e des Sciences Semlalia, Universit\'e Cadi Ayyad, LPHEA-Marrakech;$^{(d)}$LPMR, Facult\'e des Sciences, Universit\'e Mohamed Premier, Oujda;$^{(e)}$Facult\'e des sciences, Universit\'e Mohammed V, Rabat;$^{(f)}$Institute of Applied Physics, Mohammed VI Polytechnic University, Ben Guerir; Morocco.\\
% ...
$^{45}$Physics Department, Southern Methodist University, Dallas TX; United States of America.\\
% ...
$^{49}$Fakult\"{a}t Physik, Technische Universit{\"a}t Dortmund, Dortmund; Germany.\\
% ...
$^{55}$II. Physikalisches Institut, Georg-August-Universit\"{a}t G\"ottingen, G\"ottingen; Germany.\\
% ...
$^{84}$$^{(a)}$Khalifa University of Science and Technology, Abu Dhabi;$^{(b)}$New York University Abu Dhabi, Abu Dhabi;$^{(c)}$United Arab Emirates University, Al Ain;$^{(d)}$University of Sharjah, Sharjah; United Arab Emirates.\\
% ...
$^{103}$CPPM, Aix-Marseille Universit\'e, CNRS/IN2P3, Marseille; France.\\
% ...
$^{121}$Homer L. Dodge Department of Physics and Astronomy, University of Oklahoma, Norman OK; United States of America.\\
% ...
$^{155}$Raymond and Beverly Sackler School of Physics and Astronomy, Tel Aviv University, Tel Aviv; Israel.\\
% ...
$^{a}$ Also at Affiliated with an institute formerly covered by a cooperation agreement with CERN.\\
$^{b}$ Also at An-Najah National University, Nablus; Palestine.\\
$^{c}$ Also at Borough of Manhattan Community College, City University of New York, New York NY; United States of America.\\
$^{d}$ Also at Center for Interdisciplinary Research and Innovation (CIRI-AUTH), Thessaloniki; Greece.\\
$^{e}$ Also at Centre of Physics of the Universities of Minho and Porto (CF-UM-UP); Portugal.\\
$^{f}$ Also at CERN, Geneva; Switzerland.\\
% ...

\end{flushleft}


%

%
%

%

%
%
%
%

\end{document}